\documentclass{emulateapj}
\usepackage{amsmath,amssymb}

\shorttitle{A new code for numerical simulation of MHD astrophysical flows with chemistry}
\shortauthors{Kulikov I., Chernykh I., Protasov V.}

\begin{document}
\title{A new code for numerical simulation of MHD astrophysical flows with chemistry}

\author{Igor Kulikov\altaffilmark{1}, Igor Chernykh\altaffilmark{2}, Viktor Protasov\altaffilmark{3}}
\affil{Institute of Computational Mathematics and Mathematical Geophysics SB RAS, \\ 
Russia, Novosibirsk, 630090}

\altaffiltext{1}{Ph. D., Research scientist, Department of parallel algorithms of large-scale problems solving, Institute of Computational Mathematics and Mathematical Geophysics SB RAS, 630090, Novosibirsk, Russia; Research Scientist, Novosibirsk State University, 630090, Novosibirsk, Russia; Associate Professor, Novosibirsk State Technical University, 630073, Novosibirsk, Russia; kulikov@ssd.sscc.ru}

\altaffiltext{2}{Ph. D., Senior Research scientist, Siberian Supercomputer Center, Institute of Computational Mathematics and Mathematical Geophysics SB RAS, 630090, Novosibirsk, Russia; chernykh@parbz.sscc.ru}

\altaffiltext{3}{PhD Student, Novosibirsk State Technical University, 630073, Novosibirsk, Russia; 
inc\_13@mail.ru}

\begin{abstract}
The new code for numerical simulation of magnetic hydrodynamical astrophysical flows with consideration of chemical reactions is given in the paper. At the heart of the code -- the new original low-dissipation numerical method based on a combination of operator splitting approach and piecewise-parabolic method on the local stencil. The details of the numerical method are described; the main tests and the scheme of parallel implementation are shown. The chemodynamics of the hydrogen while the turbulent formation of molecular clouds is modeled.\end{abstract}


\section{Introduction}
Magnetic field plays a key role in formation and dynamics of astrophysical objects. Thus, on cosmological scales the influence of weak magnetic field, with order $\mu$G, on the dynamics of hydrodynamical instabilities and ram-pressure mechanism in galactic clusters \citep{Bruggen_2013} was studied; primarily radial orientation of the magnetic field in Virgo cluster outside the central area was defined \citep{Pfrommer_2010}, and comparison of magnetic field with radio observation was carried out \citep{Xu_2012}. The structure of magnetic field in spiral arms of the M51 galaxy was investigated \citep{Fletcher_2011}, and the evolution of the disk galaxy with consideration of influence of magnetic field was modeled \citep{Pakmor_2013}. The consideration of the influence of magnetic field plays an important role in the evolution of interstellar turbulent flows where the magnetic field is sufficiently strong \citep{Perez_2010,Mason_2011}. In the problems of evolution of MHD instabilities the power spectrum \citep{Beresnyak_2014}, sub-alfvenic flows \citep{McKee_2010}, and starburst rate \citep{Federrath_2012} were studied, and the comparison of different codes on the problem of supersonic turbulence was made \citep{Kritsuk_2011}. In the problems connected with the stellar wind, the MHD simulation is necessary too. Thus, the turbulence in stellar wind was investigated \citep{Galtier_2007}, one-dimensional MHD model of interaction of stellar wind with 67P/Churyumov-Gerasimenko comet \citep{Mendis_2014}, Halley comet \citep{Ogino_1988}, and also gas planet \citep{Johnstone_2015,Shematovich_2014} was built. In addition, the similar problems of interaction between stellar wind and stars \citep{Villaver_2012} are worth noting.

Besides earlier classical (AMR and SPH) approaches (see the methods and codes overview in \citep{Kulikov_2014,Kulikov_2015} and in classical monography \citep{Kulikovsky_2001}) a lot of new original numerical methods and codes for astrophysical MHD flows simulations were made for the last decade. For example, in CosmoMHD code \citep{Li_2008} based on TVD-ES approach MHD equations are solved in extended form with additional equations for internal energy and entropy. Such approach allows better simulations of flows with large Mach numbers \citep{Balsara_1999} because of entropy conservation law. However, the possibility of growth of entropy within shock waves is an opened issue, because the equation for entropy is formulated as inequality \citep{Godunov_2014} and, in fact, doesn't use in computations so as an equation for internal energy, except, perhaps, the areas with low-density \citep{Vshivkov_2011b}. The GOEMHD3 \citep{Skala_2015} code, based on the combination of a leap-frog, Lax and DuFort-Frankel finite-differential schemes for a non-conservative form of MHD equations, was developed for simulation of MHD flows with large Reynolds number. Nevertheless, such formulation had allowed reproducing solution with Reynolds number $10^{10}$ good enough. Also, codes based on Godunov type solvers with high order accuracy such as Athena \citep{Stone_2008}, Fish \citep{Kappeli_2011}, MPI-AMRVAC \citep{Porth_2014}, Pluto \citep{Mignone_2011}, and code based on piecewise-parabolic method on local stencil \citep{Ustyugov_2007,Ustyugov_2008} are worth noting. TVD reconstruction of numerical solution is used in all of them that is naturally in methods with high order accuracy. Besides, the Flux-CT scheme \citep{Balsara_1999b} based on Stokes theorem is used to to comply with the term  $\bigtriangledown \times B = 0$, that is more efficient than projection schemes \citep{Brackbill_1980} used in couple of codes \citep{Springel_2010}.

The new low-dissipation numerical scheme for solving equations of magnetic gas dynamics with consideration of chemical processes and its software implementation are presented in the paper. The numerical method is based on hybrid method developed earlier with combination of operator splitting approach and Godunov method in its basis  \citep{Tutukov_2011,Vshivkov_2009,Vshivkov_2011a,Vshivkov_2011b,
Godunov_2014,Kulikov_2014,Kulikov_2015,Chernykh_2015,Kulikov_2015_book,Birds_2016,
Protasov_2015,Protasov_2016}. The piecewise-parabolic method on local stencil is used in all stages of the scheme to get low dissipation of the solution. The special algorithm for building the local parabola had allowed us to fully eliminate the using of TVD reconstructions of the numerical solution in the region of discontinuous solutions. We specifically do not claim the new method as a method of high order accuracy, because this term is not fully formulated in a case of discontinuous solutions \citep{Godunov_2011}. In the first section, the numerical method is defined, and its software implementation is briefly described. In the second section, the one- and two-dimensional tests are shown. The third section is devoted to simulation of 3D MHD flows.

\section{Numerical method description}
In this paper, we will only consider the MHD flows, and the model problems will be considered just in MHD approximation. Thus, the system of equations of gravitational multicomponent magnetic gas dynamics in 3D cartesian coordinates taking into account the function of heating and cooling is used:
$$
\frac{\partial \rho}{\partial t} + \bigtriangledown \cdot \left( \rho \mathbf{v} \right) = 0
$$
$$
\frac{\partial \rho_i}{\partial t} + \bigtriangledown \cdot \left( \rho_i \mathbf{v} \right) = \mathcal{S}_i
$$
$$
\frac{\partial \rho \mathbf{v}}{\partial t} + \bigtriangledown \cdot \left( \rho \mathbf{v} \mathbf{v} - \mathbf{B} \mathbf{B} \right) = - \bigtriangledown p^{*} - \rho \bigtriangledown \Phi
$$
$$
\frac{\partial \rho E}{\partial t} + \bigtriangledown \cdot \left( \left( \rho E + p^{*} \right)  \mathbf{v} - \mathbf{B} \left( \mathbf{B} \cdot \mathbf{v} \right) \right) = - \left( \rho \mathbf{v} \cdot \bigtriangledown \Phi \right) + \Gamma - \Lambda
$$
$$
\frac{\partial \rho \varepsilon}{\partial t} + \bigtriangledown \cdot \left( \rho \varepsilon \mathbf{v} \right) = - \left( \gamma - 1 \right) \rho \varepsilon \bigtriangledown \cdot \left( \mathbf{v} \right) + \Gamma - \Lambda 
$$
$$
\frac{\partial \mathbf{B}}{\partial t} = \bigtriangledown \times \left( \mathbf{v} \times \mathbf{B}  \right)
$$
$$
\bigtriangleup \Phi = 4 \pi G \rho
$$
the condition of non-divergency of magneic field
$$
\bigtriangledown \cdot \left( \mathbf{B} \right) = 0
$$
where $\rho = \sum_{i} \rho_i$ -- density, $\rho_i$ -- density of each component of the gaseous mixture, $\mathcal{S}_i$ -- formation rate of $i$-th component of the mixture, $\mathbf{v}$ -- velocity, $\mathbf{B}$ -- magnetic field, $p = \rho \varepsilon (\gamma - 1)$ -- pressure, $\rho \varepsilon$ -- internal energy, $p^{*} = p + \mathbf{B}^2 / 2$ -- full pressure, $\gamma$ -- adiabatic index, $\rho E = \rho \varepsilon + \rho \mathbf{v}^2 / 2 + \mathbf{B}^2 / 2$ -- full mechanical energy, $\Phi$ -- gravitational potential, $\Gamma$ -- heating function, $\Lambda$ -- cooling function, $G$ -- gravitational constant.

The method of solving equations of gravitational multicomponent magnetic gas dynamics is based on a combination of operator splitting approach and Godunov method with using the piecewise-parabolic method on the local stencil. It consists of the following stages: 
\begin{enumerate}
\item eulerian stage, at wich the equations for density, impulse, full and internal energy are solved without consideration of advective terms and functions of heating and cooling, but with consideration of work of gravitational force;
\item recomputation of magnetic field with conservation of condition $\bigtriangledown \cdot \left( \mathbf{B} \right) = 0$ with using of Flux-CT scheme;
\item lagrangian stage, at which the advection of density, momentum, full, and internal energy happens;
\item solving of homogeneous differential equations in each cell of computational domain to compute concentration of gas mixture;
\item consideration of subcell processes of cooling/heating;
\item regularization of numerical solution;
\item solving of Poisson equation to compute gravitational potential.
\end{enumerate} 
Before proceeding to the detailed description of each stage let us describe two procedures, on which the eulerian and lagrangian stages are based -- the procedure of building the local parabolas, which will be used in solution of the Riemann problem at each stage; and the procedure of using the fourth-order Runge-Kutta method that used on the eulerian and lagrangian stages separately.

\subsection{Procedure of building of the local parabolas}
For definitness we will constuct piecewise-parabolic function of particular parameter $q(x)$ on the regular grid with step $h$ in the interval $[x_{i-1/2},x_{i+1/2}]$. In general, parabola could be written as:

$$
q(x) = q_{i}^{L} + \xi \left( \bigtriangleup q_{i} + q_{i}^{(6)} (1 - \xi) \right)
$$
where $q_{i}$ -- value in the center of cell, $\xi = (x - x_{i-1/2})h^{-1}$, $\bigtriangleup q_{i} = q_{i}^{L} - q_{i}^{R}$ and $q_{i}^{(6)} = 6 (q_{i} - 1/2 (q_{i}^{L} + q_{i}^{R}))$ while maintaining conservatism, that is:
$$
q_{i} = h^{-1} \int_{x_{i-1/2}}^{x_{i+1/2}} q(x) dx
$$
Let us give the detailed procedure of building of the parabola and parameters $q_{i}^{R}$, $q_{i}^{L}$, 
$\bigtriangleup q_{i}$, $q_{i}^{6}$. To constuct the values $q_{i}^{R} = q_{i+1}^{L} = q_{i+1/2}$ fourth-order interpolational function will be used:
$$
q_{i+1/2} = 1/2(q_{i} + q_{i+1}) - 1/6 (\delta q_{i+1} - \delta q_{i})
$$
where $\delta q_{i}  = 1/2 (q_{i+1} - q_{i-1})$. Further, we describe the algorithm of building the local parabola. The input is the values in centers of cells $q_{i}$. The output of the algorithm is all of the parameters of piecewise-parabolic functions in every interval $[x_{i-1/2},x_{i+1/2}]$.

\textbf{Step 1.} At the first ste the values $\delta q_{i}  = 1/2 (q_{i+1} - q_{i-1})$ are constructed. To do this we need to know only nearby cells $q_{i+1}, q_{i-1}$. To eliminate the extrema of functions the modification of the last formula for $\delta q_{i}$ is used as follows:
$$
\delta_{m} q_{i} = \left\lbrace \begin{array}{c}
min(\vert \delta q_{i} \vert, 2 \vert q_{i+1} - q_{i} \vert, 2 \vert q_{i} - q_{i-1} \vert) sign(\delta q_{i}), \\
\qquad (q_{i+1} - q_{i})(q_{i} - q_{i-1}) > 0 \\
0, (q_{i+1} - q_{i})(q_{i} - q_{i-1}) \leq 0
\end{array}  \right.
$$
The exchange of the one layer of overlapping should be done with using of MPI in the case of a parallel implementation on the architectures with distributed memory. Then, the values on the borders are recomputed with using of the fourth-order interpolant:$$
q_{i}^{R} = q_{i+1}^{L} = q_{i+1/2} = 1/2(q_{i} + q_{i+1}) - 1/6 (\delta_{m} q_{i+1} - \delta_{m} q_{i})
$$

\textbf{Step 2.} At the second step of the algorithm the local parabola is constructed with using of the following formula:
$$
\bigtriangleup q_{i} = q_{i}^{L} - q_{i}^{R}
\qquad
q_{i}^{(6)} = 6 (q_{i} - 1/2 (q_{i}^{L} + q_{i}^{R}))
$$
The values on the borders $q_{i}^{L}, q_{i}^{R}$ in case of non-monotonic local parabola (it could happen in dicontinuities) are reconstructed according to the formulas:
$$
q_{i}^{L} = q_{i}, q_{i}^{R} = q_{i}, (q_{i}^{L} - q_{i})(q_{i} - q_{i}^{R}) \leq 0
$$
$$
q_{i}^{L} = 3q_{i} - 2q_{i}^{R},  \bigtriangleup q_{i} q_{i}^{(6)} > (\bigtriangleup q_{i})^{2}
$$
$$
q_{i}^{R} = 3q_{i} - 2q_{i}^{L}, \bigtriangleup q_{i} q_{i}^{(6)} < - (\bigtriangleup q_{i})^{2}
$$
Thus, the boundary values satisfy the conditions of monotonicity.

\textbf{Step 3.} At the third step the parabola parameters are reconstructed with consideration of new values in boundary cells:
$$
\bigtriangleup q_{i} = q_{i}^{L} - q_{i}^{R}
$$
$$
q_{i}^{(6)} = 6 (q_{i} - 1/2 (q_{i}^{L} + q_{i}^{R}))
$$
It is worth noting, that parabolas could have a discontinuity on the borders of cells, that leads to the need of solving of the Riemann problem for parabolas in case of using of classical piecewise-parabolic method (PPM). In our case the local parabolas are used as a part of the Riemann problem.

\textbf{Step 4.} At the fourth step additional monotonization of parabola is done. If we are in the region of discontinuity of the function, then the additional amendments are made:
$$
q_{i}^{L,+} = q_{i} - \frac{1}{4} \delta_{m} q_{i} \qquad q_{i}^{R,+} = q_{i} + \frac{1}{4} \delta_{m} q_{i}
$$
Additional criteria is introduced:
$$
\eta = - h^2 \frac{\delta_{m}^2 q_{i+1} - \delta_{m}^2 q_{i-1}}{q_{i+1} - q_{i-1}}
$$
If one of the follow inf conditions is satisfied:
$$
\vert q_{i+1} - q_{i-1} \vert - \frac{\min (\vert q_{i+1} \vert, \vert q_{i-1} \vert, \vert q_{i+1} \vert + \vert q_{i-1} \vert)}{100}  \leq 0
$$
$$
q_{i+1} q_{i-1} > 0 
$$
the value of the criteria $\eta$ is set to zero. The weight of values $q_{i}^{L,+}$ и $q_{i}^{R,+}$ in the computational scheme is defined by the formula:
$$
\hslash = \max(\min(20(\eta-0.05),1),0) 
$$
Final values of the flows are computed by the formulas:
$$
q_{i}^{L,FINAL} = (1 - \hslash)q_{i}^{L,+} + \hslash q_{i}^{L} 
$$
$$
q_{i}^{R,FINAL} = (1 - \hslash)q_{i}^{R,+} + \hslash q_{i}^{R}
$$
The last two values are used to determ the quantities $q_{i}^{L}$ и $q_{i}^{R}$. Such additional monotonicity is done for all magneto-hydrodynamics quantities in contrast to classical procedure in \citep{Collela_1984}, and also slightly different ways to compute the gradient of the solution were experimentally found.

\textbf{Step 5.} At the fifth step the final reconstruction of the parabola with consideration of new values on the borders of the cell is made:
$$
\bigtriangleup q_{i} = q_{i}^{L} - q_{i}^{R}
$$
$$
q_{i}^{(6)} = 6 (q_{i} - 1/2 (q_{i}^{L} + q_{i}^{R}))
$$
As a result the local parabola in every cell $[x_{i-1/2},x_{i+1/2}]$ is computed. Notice, that monotonicity of the numerical solution have a place only at the stage of building of the local parabola, which is used for solving the Riemann problem at each stage.

\subsection{Runge-Kutta time integration scheme}
Using the finite-volume approximation of eulerian and lagrangian stages, described further, the numerical scheme could be written in form of ordinary differential equation as follows:
$$
\frac{d \mathcal{Q}}{dt} = \mathcal{R}
$$
where $\mathcal{Q}$ -- magnetic-hydrodynamics parameters, $\mathcal{R}$ -- finite-volume approximation of each stage. In this case the Runge-Kutta scheme for approximation of the derivations by time will be used to compute the solution at each stage:
$$
\mathcal{Q}^{\left(n+1/3\right)} = \mathcal{Q}^{\left(n\right)} + \tau \mathcal{R}^{\left(n\right)}
$$
$$
\mathcal{Q}^{\left(n+2/3\right)} = \frac{3}{4}\mathcal{Q}^{\left(n\right)} + \frac{1}{4}\mathcal{Q}^{\left(n+1/3\right)} + \frac{\tau}{4} \mathcal{R}^{\left(n+1/3\right)}
$$
$$
\mathcal{Q}^{\left(n+1\right)} = \frac{1}{3}\mathcal{Q}^{\left(n\right)} + \frac{2}{3}\mathcal{Q}^{\left(n+2/3\right)} + \frac{2\tau}{3} \mathcal{R}^{\left(n+2/3\right)}
$$
where $\mathcal{Q}^{\left(i\right)}$ -- the solution at each stage on the $i$-th layer of time, $\mathcal{R}^{\left(i\right)}$ -- finite-volume approximation of equations on the $i$-th layer of time.

\subsection{Compliance with the Courant–Friedrichs–Lewy condition}
To choose the time step $\tau$ the speed of sound $c = \sqrt{\frac{\gamma p}{\rho}}$, alfven speed of sound $c_a = \vert \frac{B_x}{\sqrt{\rho}} \vert$, fast $c_f$ and slow $c_s$ magnetic speeds:
$$
c_f = \sqrt{\frac{\left( c^2 + b^2 \right) + \sqrt{\left( c^2 + b^2 \right)^2 - 4 c^2 c_a^2}}{2}}
$$
$$
c_s = \sqrt{\frac{\left( c^2 + b^2 \right) - \sqrt{\left( c^2 + b^2 \right)^2 - 4 c^2 c_a^2}}{2}}
$$
where $b = \sqrt{B_x^2 + B_y^2 + B_z^2}$ are determined in each cell. Then the time step is computed according to the equation:
$$
\tau = \min\left( \frac{CFL \times h}{v + b + c + c_a + c_s + c_f}  \right)
$$
where $v = \sqrt{v_x^2 + v_y^2 + v_z^2}$ -- speed, $h$ -- lenght of the edge of a cell, $CFL = 0.2$ -- the Courant–Friedrichs–Lewy number.

\subsection{Eulerian stage}
At the eulerian stage of the scheme the linearized Godunov method is used. Gravitational force is calculated with using of central differential scheme because of smoothness of gravitational potential. The Riemann problem with piecewise-parabolic initial conditions in all directions is formulated to compute magnetic hydrodynamic flows on the borders of each computational cell:
$$
\frac{\partial q}{\partial t} + \mathcal{B} \frac{\partial q}{\partial x} = 0
$$
In case of MHD equations for eulerian stage the vector $q = \left( v_x, v_y, v_z, B_y, B_z, p \right)^T$, and the matrix $\mathcal{B}$ on each border of cells is written as follows: 
$$
\mathcal{B} = \begin{pmatrix}
 0 & 0 & 0 & \frac{B_y}{\rho} & \frac{B_z}{\rho} & \frac{1}{\rho} \\  
 0 & 0 & 0 & -\frac{B_x}{\rho} & 0 & 0 \\  
 0 & 0 & 0 & 0 & -\frac{B_x}{\rho} & 0 \\  
 B_y & -B_x & 0 & 0 & 0 & 0 \\  
 B_z & 0 & -B_x & 0 & 0 & 0 \\  
 \gamma p & 0 & 0 & 0 & 0 & 0 
\end{pmatrix} 
$$
To average magnetic-hydrodynamics values from left (L) and right (R) cells on the border between them the following equations are used: 
$$
\rho = \frac{\sqrt{\rho^L}\rho^L + \sqrt{\rho^R}\rho^R}{\sqrt{\rho^L} + \sqrt{\rho^R}}
$$
$$
v_{[x,y,z]} = \frac{\sqrt{\rho^L}v_{[x,y,z]}^L + \sqrt{\rho^R}v_{[x,y,z]}^R}{\sqrt{\rho^L} + \sqrt{\rho^R}}
$$
$$
B_{[x,y,z]} = \frac{\sqrt{\rho^L}B_{[x,y,z]}^R + \sqrt{\rho^R}B_{[x,y,z]}^L}{\sqrt{\rho^L} + \sqrt{\rho^R}}
$$
after that the speed of sound $c$, alfvenic speed of sound $c_a$, fast $c_f$ and slow $c_s$ magnetic speed could be calculated by the equations form the previous section. For the eigenvalue-decomposition matrix $\mathcal{B}$ the definition of the following parameters should be extended:
$$
\left( \alpha_f, \alpha_s \right) = 
\left\{
\begin{array}{@{\,}c@{\quad}l@{}}
\frac{ \left( \sqrt{c^2 - c_s^2}, \sqrt{c_f^2 - c^2} \right) }{\sqrt{c_f^2 - c_s^2}} & B_y^2 + B_z^2 > 0, \gamma p \neq B_x^2 \\ 
\left( \frac{1}{\sqrt{2}}, \frac{1}{\sqrt{2}} \right) & B_y^2 + B_z^2 = 0, \gamma p = B_x^2
\end{array}\right.
$$ 
$$
\left( \beta_f, \beta_s \right) = 
\left\{
\begin{array}{@{\,}c@{\quad}l@{}}
\frac{ \left( B_y, B_z \right) }{\sqrt{B_y^2 + B_z^2}} & B_y^2 + B_z^2 > 0 \\ 
\left( \frac{1}{\sqrt{2}}, \frac{1}{\sqrt{2}} \right) & B_y^2 + B_z^2 = 0 \end{array}\right.
$$
The matrix $\mathcal{B}$ could be written as eigenvalue-decomposition $\mathcal{B} = R \Omega L$, where $R$ and $L$ are mutually orthogonal matrices $RL = LR = I$ of right and left eigenvectors (their form is given in appendix), $\Omega$ is a diagonal matrix with eigenvalues:
$$
\lambda_1 = c_f \qquad \lambda_2 = -c_f
\quad
\lambda_3 = c_s \qquad \lambda_4 = -c_s
$$
$$
\lambda_5 = c_a \qquad \lambda_6 = -c_a
$$
Replacing $s = Lq$ we get the system
$$
\frac{\partial s}{\partial t} + \Omega \frac{\partial s}{\partial x} = 0
$$
which could be solved analytically but with the consideration that initial conditions $s^0$ for this problem is a piecewise-parabolic functions. Thus, the solution of Riemann problem for the last system of equations could be formulated in form:
$$
\mathbf{s}_1 = s_1^0 \left( - c_f \tau \right) \qquad \mathbf{s}_2 = s_2^0 \left( c_f \tau \right)
\qquad \mathbf{s}_3 = s_3^0 \left( - c_s \tau \right) 
$$
$$
\mathbf{s}_4 = s_4^0 \left( c_s \tau \right)
\qquad
\mathbf{s}_5 = s_5^0 \left( - c_a \tau \right) \qquad \mathbf{s}_6 = s_6^0 \left( c_a \tau \right)
$$
Depending on the sign of eigenvalue the integration should be made by the left or the right parabola. Using the notations from section (2.1) solution could be written as follows:
$$
q(- \nu t) = q_{i}^{R} - \frac{\nu t}{2h} \left( \bigtriangleup q_{i} - q_{i}^{6} \left(1 - \frac{2 \nu t}{3h} \right) \right)
$$
$$
q(\nu t) = q_{i}^{L} + \frac{\nu t}{2h} \left( \bigtriangleup q_{i} + q_{i}^{6} \left(1 - \frac{2 \nu t}{3h} \right) \right)
$$
where $\nu$ -- modulus of the eigenvalue, $i$ -- the number of cell depending on the consideration of the left or the right parabola. After the Riemann problem is solved for the vector $s$, using the replacement $q = Rs$ the solution of the Riemann problem is calculated $\mathbf{V_{x}}, \mathbf{V_{y}}, \mathbf{V_{z}}, \mathbf{B_{y}}, \mathbf{B_{z}}, \mathbf{P}$, that is used in the finite-volume approximation further (their final form is given in appendix).

\subsection{Satisfying the condition $\bigtriangledown \cdot \left( \mathbf{B} \right) = 0$}
To satisfy the condition $\bigtriangledown \cdot \left( \mathbf{B} \right) = 0$ the Flux-CT scheme \citep{Balsara_1999b} based on Stokes theorem was used:
$$
\frac{\partial \mathbf{B}}{\partial t} = \bigtriangledown \times \left( \mathbf{v} \times \mathbf{B}  \right)
$$
\begin{figure}
\begin{center}
\includegraphics[width=0.5\linewidth]{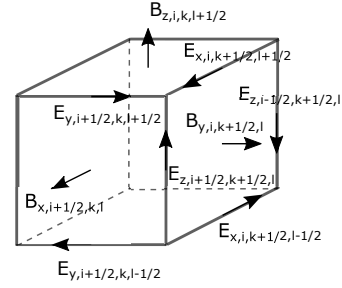}
\end{center}
\caption{The schemas of cell $(i,k,l)$}
\label{divbzeroscheme}
\end{figure}
We will use follow schemas of cell $(i,k,l)$ (see. fig. \ref{divbzeroscheme}) and condition $h_x = h_y = h_z = h$. The magnetic filed vector will define on border of cells:
$$
B_{x,i+1/2,k,l}^{n+1} = B_{x,i+1/2,k,l}^{n} 
$$
$$
- \frac{\tau}{h} \left( E_{z,i+1/2,k+1/2,l} - E_{z,i+1/2,k-1/2,l} \right)
$$
$$
+ \frac{\tau}{h} \left( E_{y,i+1/2,k,l+1/2} - E_{y,i+1/2,k,l-1/2} \right)
$$
$$
B_{y,i,k+1/2,l}^{n+1} = B_{y,i,k+1/2,l}^{n} 
$$
$$
- \frac{\tau}{h} \left( E_{x,i,k+1/2,l+1/2} - E_{x,i,k+1/2,l-1/2} \right)
$$
$$
+ \frac{\tau}{h} \left( E_{z,i+1/2,k+1/2,l} - E_{z,i-1/2,k+1/2,l} \right)
$$
$$
B_{z,i,k,l+1/2}^{n+1} = B_{z,i,k,l+1/2}^{n} 
$$
$$
- \frac{\tau}{h} \left( E_{y,i+1/2,k,l+1/2} - E_{y,i-1/2,k,l+1/2} \right)
$$
$$
+ \frac{\tau}{h} \left( E_{x,i,k+1/2,l+1/2} - E_{x,i,k-1/2,l+1/2} \right)
$$
The vector of electricity field $\vec{E} = - \vec{u} \times \vec{B} = \vec{B} \times \vec{u}$ define in follow form:
$$
E_x = B_y u_z - B_z u_y, E_y = B_z u_x - B_x u_z, E_z = B_x u_y - B_y u_x
$$
In last formulas we will use solution of Riemann problems (details equations can found in appendix).
On next step the vector of magnetic field will projected to center of cells by means follow equation:
$$
B_{x,ikl} = \frac{1}{2} \left( B_{x,i+1/2,k,l} + B_{x,i-1/2,k,l} \right)
$$
$$
B_{y,ikl} = \frac{1}{2} \left( B_{y,i,k+1/2,l} + B_{y,i,k-1/2,l} \right)
$$
$$
B_{z,ikl} = \frac{1}{2} \left( B_{z,i,k,l+1/2} + B_{z,i,k,l-1/2} \right)
$$

\subsection{Lagrangian stage}
At the lagrangian stage the advection of hydrodynamical parameters is carried out and the equations at this stage are the following:
$$
\frac{\partial q}{\partial t} + \nabla \cdot (q \mathbf{v}) = 0
$$
where $q$ could be density $\rho$, momentum $\rho \mathbf{v}$, density of the full mechanical $\rho E$ or internal $\rho \epsilon$ energy of the gas. To solve the equations we use the similar approach that used at the eulerian stage. To compute the flow $F = q \mathbf{v}$ при $\lambda = \vert \mathbf{v} \vert$ following formula is used:
$$
F = \mathbf{v} \times \left\lbrace
\begin{array}{c}
q\left(- \lambda \tau \right), \mathbf{v} \geq 0 \\
q\left(\lambda \tau \right), \mathbf{v} < 0
\end{array}
\right.
\qquad
\mathbf{v} = \frac{ \mathbf{v}_{L}\sqrt{\rho_{L}} + \mathbf{v}_{R}\sqrt{\rho_{R}} }{ \sqrt{\rho_{L}} + \sqrt{\rho_{R}} }
$$
where $q\left(- \lambda \tau \right)$ and $q\left( \lambda \tau \right)$ -- piecewise-parabolic functions for the quantity $q$. To construct the piecewise-parabolic solution the similar procedure is used.

\subsection{Chemistry}
Chemical reactions for the $i$-th component of the mixture are considered in following form:
$$
\frac{d n_i}{d t} = C_{i} \left( T, n_j \right) - D_{i} \left( T, n_j \right) n_i
$$
where $C_{i}$ -- speed of construction of $i$-th component, $D_{i}$ -- speed of destruction of $i$-th component. To solve such differential equations the scheme of inverse differentiation is used:
$$
n_i^{t+\tau} = n_i^{t} + \tau \frac{C_{i} - D_{i} n_i^{t}}{1 + \tau D_{i}}
$$
We understand that it is, possibly, not the best way and, for example, using the code KROME \citep{Grassi_2012} is more efficient but a similar approach was successfully used in works \citep{Glover_2007,Anninos_1997}.
\subsection{Subgrid physics}
To consider the subgrid physics the following equations are solved: 
$$
\frac{\partial \rho E}{\partial t} = \Gamma - \Lambda
$$
$$
\frac{\partial \rho \varepsilon}{\partial t} = \Gamma - \Lambda 
$$
in each cell with using of Euler method for solving the ODE. There is no matter to use complex way of approximation of Runge-Kutta type because the values of heating and cooling functions are constant while the time step in each cell of the computational domain.
\subsection{Regularization of numerical solution}
At the stage of regularization of the solution the correction of speed on the gas-vacuum interface, where the condition $(E - \vec{v}^{2}/2 - \mathbf{B}^2 / 2\rho)/E \geq 10^{-3}$ is satisfied, is done with using of approach similar to the one described in \citep{Vshivkov_2011b}:
$$
\vert \mathbf{v} \vert = \sqrt{2\frac{\rho E - \rho \varepsilon - \mathbf{B}^2 / 2}{\rho}}  
$$
in the rest area the correction, that guarantees non-decreasing of entropy, like in work \citep{Godunov_2014} is made:$$
\rho \epsilon = \rho E - \rho \mathbf{v}^2 / 2 - \mathbf{B}^2 / 2
$$
Such modification allows the detailed balance of energies and guarantee non-decreasing of entropy.

\subsection{Solution of the Poisson equation}
To solve the Poisson equation the 27-point pattern with the following scheme of solution in the harmonic space is used:
$$
\Phi_{jmn} = \frac{ \frac{2}{3} \pi h^2 \rho_{jmn}}
{1-\frac{3-2sin^{2}(\frac{\pi j}{I})}{3}\frac{3-2sin^{2}(\frac{\pi m}{K})}{3}\frac{3-2sin^{2}(\frac{\pi n}{L})}{3}}
$$
The Fast Fourier Transform is used to make a transition into the harmonic space, that is in finding the transition coefficients. The FFT is in a heart of the method of solving the Poisson equation/ To perform it on supercomputers with distributed memory the FFTW library \citep{FFTW_2005} was used. The library is based on procedure ALLTOALL, that ''transport'' 3D array redistributing huge amount of memory between processes. Certainly, it is expensive network operation demanding to eliminate using of the whole algorithm if we have any significant count of processors. Nevertheless, this procedure doesn't get much time while using of InfiniBand network infrastructure and, apparently, optimized in low network level \citep{Kalinkin_2009}.

\subsection{Parallel implementation}
The parallel implementation is based on the geometrical decomposition of the computational area with one layer of overlapping of the subregions by using of MPI. The study of scalability of the code was made with using of equipment of SSCC on 1 to 768 of Intel Xeon X5670 cores.
\begin{figure}
\centering
\includegraphics[width = 0.8\linewidth]{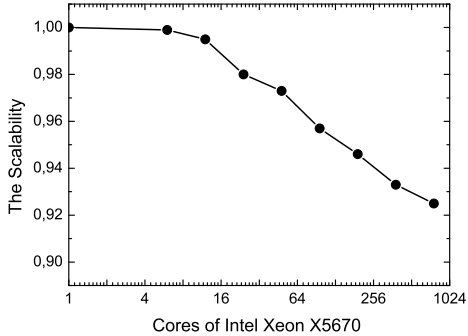}
\caption{Scalability of program implementation on the cluster NKS-30T SSCC. in numerical experiment the full MHD model was used, the grid with size $256^3$ was computed at each core.}
\label{scalability}
\end{figure}
The 93 \% efficiency was achieved on 768 computational cores (see fig. \ref{scalability}).

\section{Verification with 1D and 2D tests}
\subsection{One dimensional shock tube problem}
\begin{figure}[ht]
\centering
\includegraphics[width=0.32\linewidth]{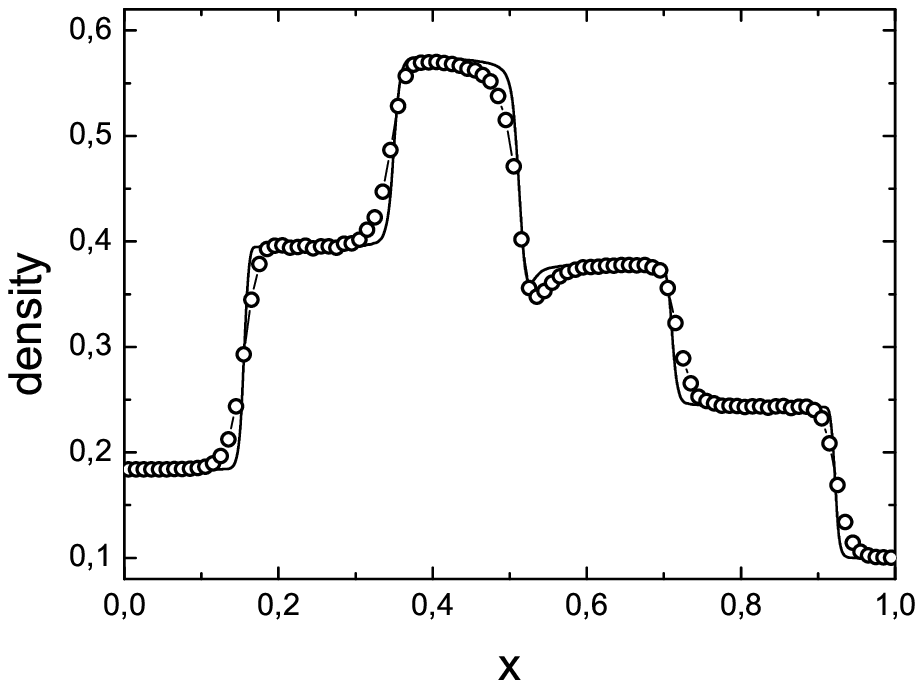}
\includegraphics[width=0.32\linewidth]{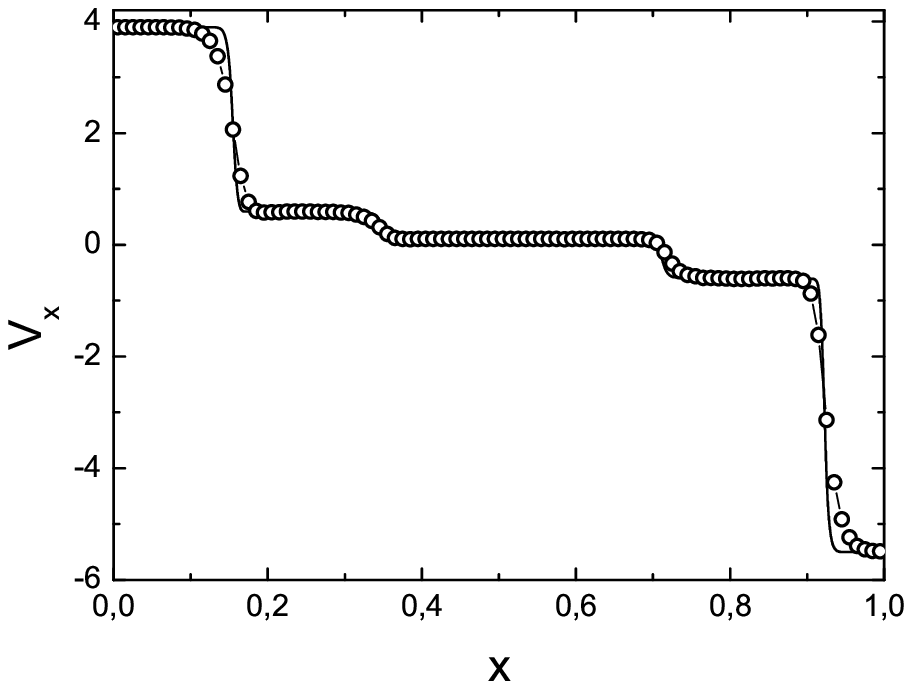} \\
\includegraphics[width=0.32\linewidth]{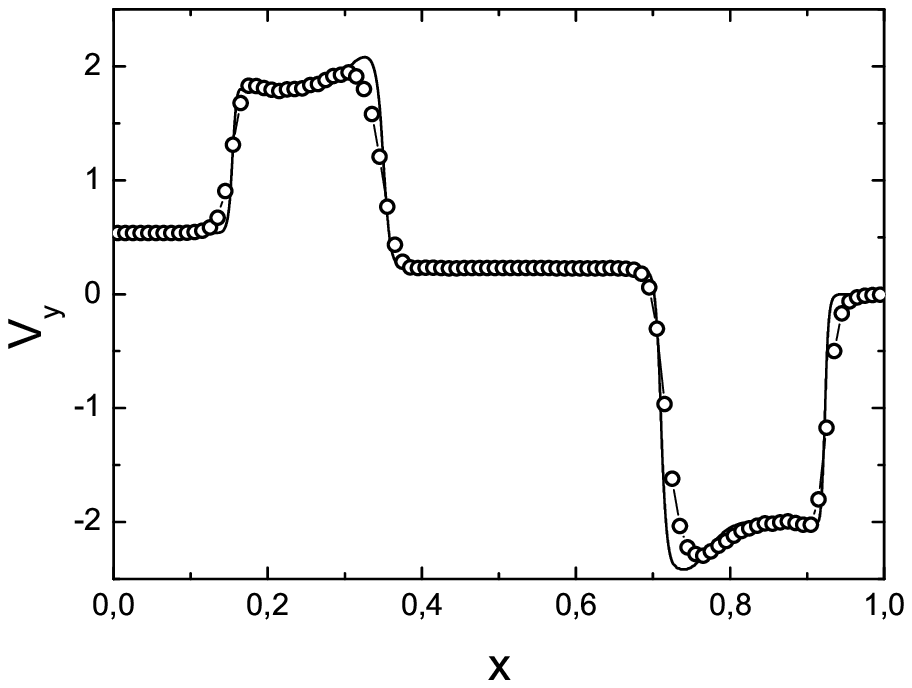}
\includegraphics[width=0.32\linewidth]{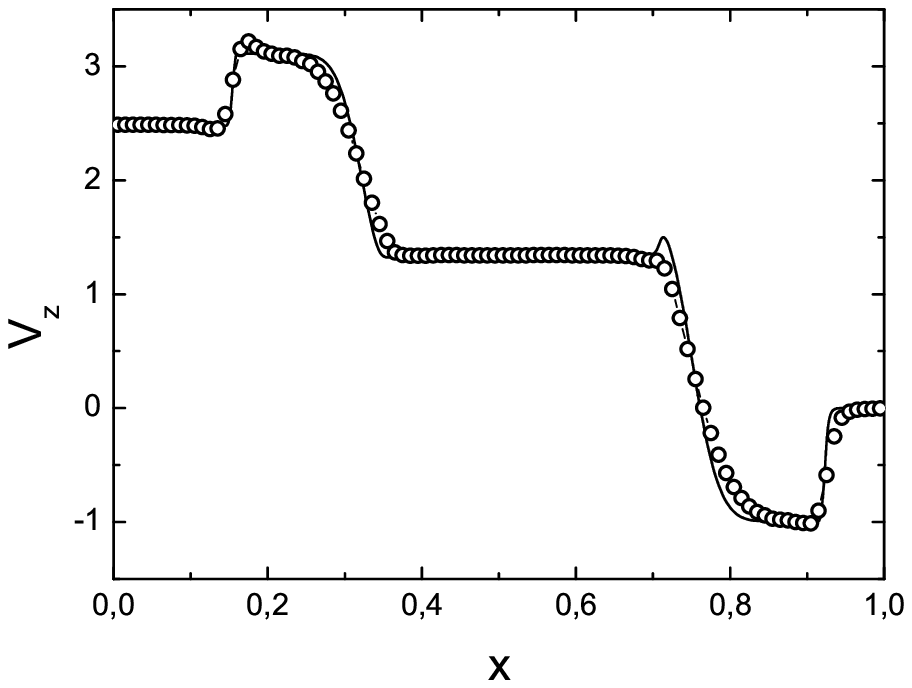} \\
\includegraphics[width=0.32\linewidth]{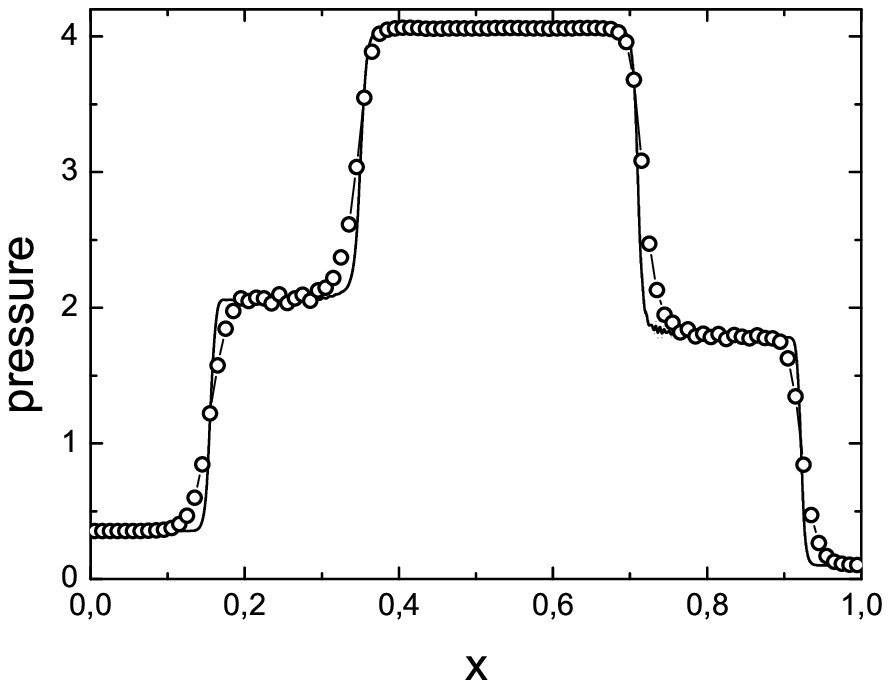}
\includegraphics[width=0.32\linewidth]{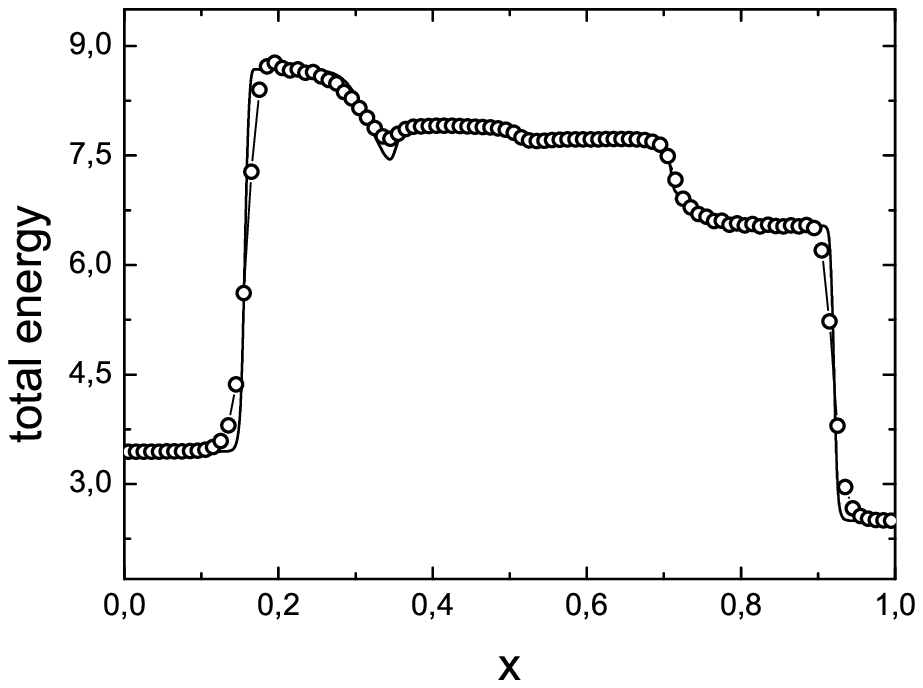} \\
\includegraphics[width=0.32\linewidth]{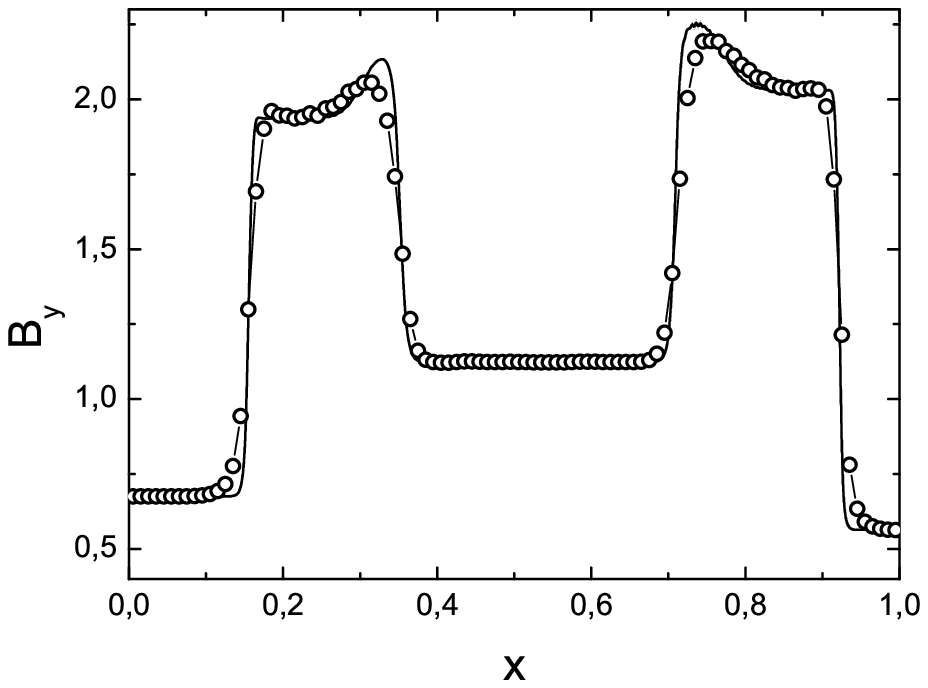}
\includegraphics[width=0.32\linewidth]{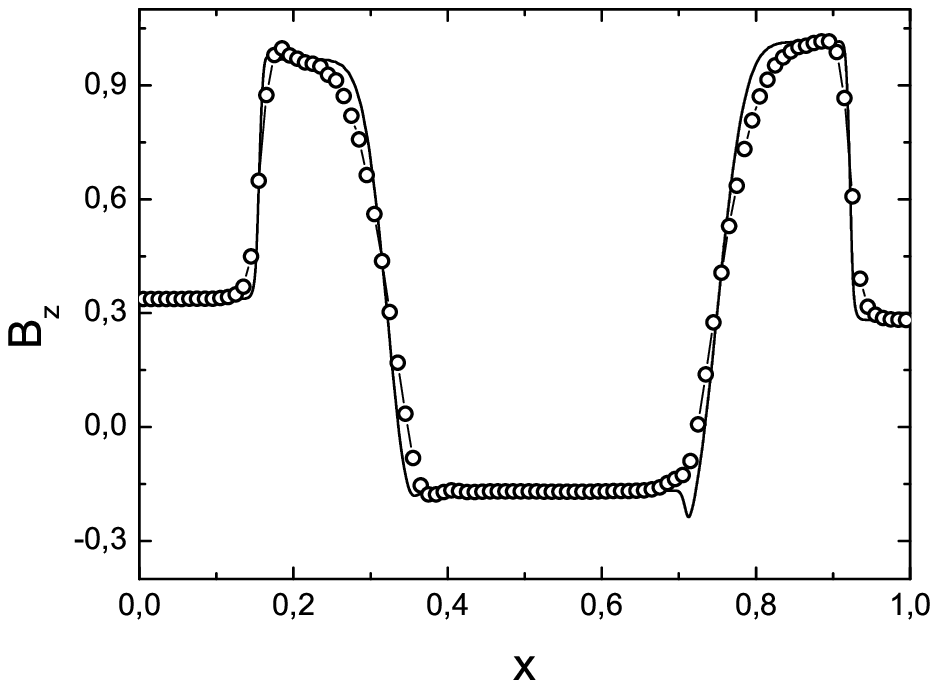} 
\caption{1D Riemann problem for MHD equations. In the figures are (from left to right, from up to down) density distribution, three components of velocity, pressure, full energy, longitude components of the magnetic field in $t = 0.15$. The solution computed with using of 100 cells is marked with symbol $\circ$; the solution computed with using of 1000 cells is shown with solid line.}
\label{pogorelov}
\end{figure}

To verify the method the 1D problem was formulated. The solution of this problem include each type of MHD shocks evolving by both regions separated by contact discontinuity \citep{Barmin_1996}. The problem is being solved in region $[0; 1]$; initial discontinuity is being set in point $x_0 = 0.5$. To the left of the discontinuity the gas parameters are $\left( \rho, p, v_x, v_y, v_z, B_y/\sqrt{4 \pi} , B_z/\sqrt{4 \pi} \right) = \left( 0.18405, 0.3541, 3.8964, 0.5361, 2.4866, 2.394, 1.197 \right)$; on the right side parameters of the gas has a more simple form $\left( \rho, p, v_x, v_y, v_z, B_y/\sqrt{4 \pi} , B_z/\sqrt{4 \pi} \right) = \left( 0.1, 0.1, -5.5, 0, 0, 2, 1 \right)$, $x$-component of magnetic field $B_x/\sqrt{4 \pi} = 4$, adiabatic index $\gamma = 1.4$. The numerical solution of the problem in the moment $t = 0.15$ is shown in figure \ref{pogorelov}.

Notice, that each shock is reproduced correctly \citep{Barmin_1996}; smearing of MHD shocks is propagated not more than on three cells; there are no oscillations of the solution near the discontinuity region, and also contact discontinuity is reproduced correctly.

\subsection{Orszag-Tang Vortex Test}
\begin{figure}[ht]
\centering
\includegraphics[width=0.5\linewidth]{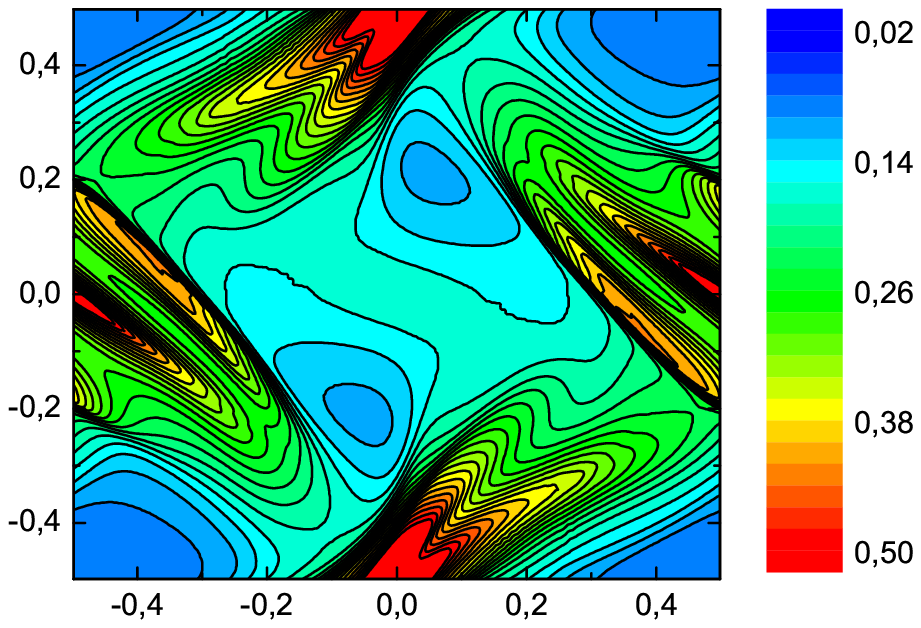} \\
\includegraphics[width=0.5\linewidth]{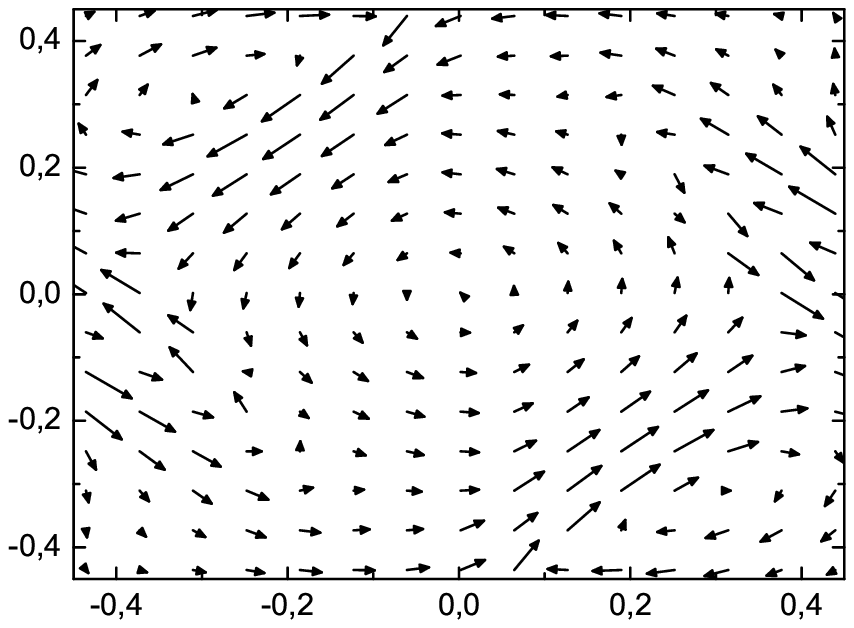} 
\caption{Orszag-Tang Vortex Test. Density distribution (top) and magnetic field (bottom) in the moment of time $t = 0.2$. In the numerical experiment was used the grid with $256^2$ cells.}
\label{orszagtang}
\end{figure}

Orszag-Tang Vortex problem \citep{Orszag_1979} is the most popular model for testing the transition to supersonic turbulence, and it verifies how correctly the code reproduce formation of shocks and their interaction. Also the condition $\bigtriangledown \cdot \left( \mathbf{B} \right) = 0$ could be tested on this problem. In the problem we consider the region $[0;1]^2$ with periodic border conditions in each direction, that is filled uniformly with density $\rho = 25/(36 \pi)$ ad pressure $p = 5/(12 \pi)$. Initial speed  $v_x = - \sin(2\pi y)$ and $v_y = \sin(2\pi x)$. Initial magnetic field $B_x = -B_0 \sin(2\pi y)$ and $B_y = B_0 \sin(4\pi x)$, where $B_0 = 1/\sqrt{4 \pi}$. Adiabatic index $\gamma = 5/3$. Numerical solution of the problem in the moment of time $t = 0.2$ is shown in figure \ref{orszagtang}.
Notice, that density field and structure of vector magnetic field are consistent with many results of other authors.

\section{Simulation of 3D MHD flows}
To verify the numerical method and program implementation in 3D, three problems in MHD statement were studied: the collision of two galaxy clusters with different mass, similar to Bullet cluster collision scenario \citep{Mastropietro_2008,Lage_2014}; the problem of interaction of molecular cloud and interstellar medium \citep{Villaver_2012}; the problem of evolution of MHD turbulence of an interstellar medium \citep{Ustyugov_2009,Kritsuk_2009} with consideration of chemical reactions \citep{Glover_2007}. The first two problems we are considering as some kind of tests and hoping to use this code to compute analogous but more complex problems. The third one was solved in full multiphase MHD statement with consideration of chemokinetics of hydrogen.

\subsection{Collision of two galaxy clusters}
\begin{figure}[ht]
\centering
\includegraphics[width=0.5\linewidth]{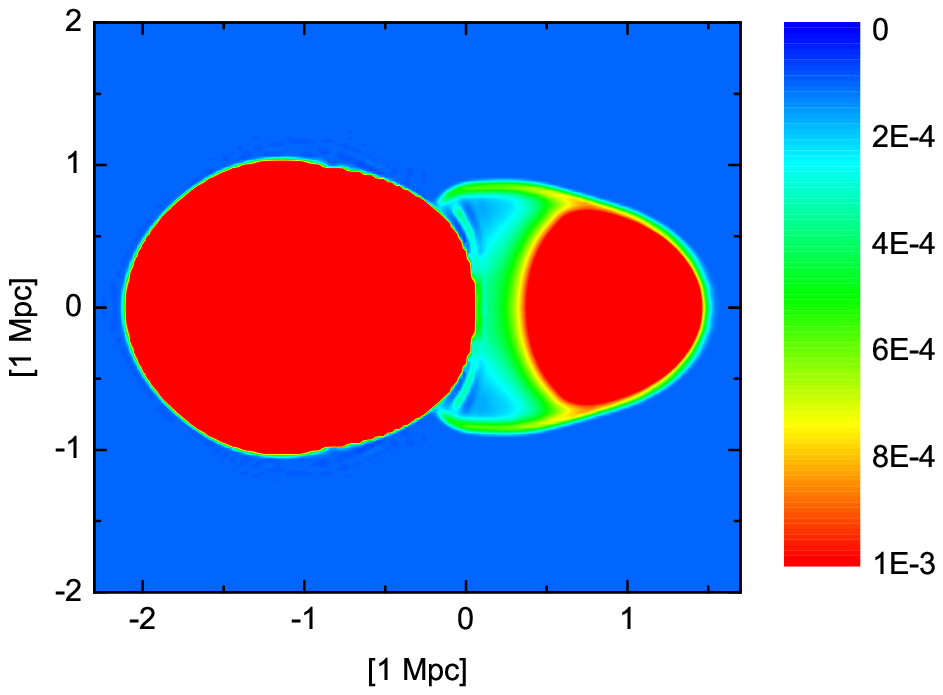} \\
\includegraphics[width=0.5\linewidth]{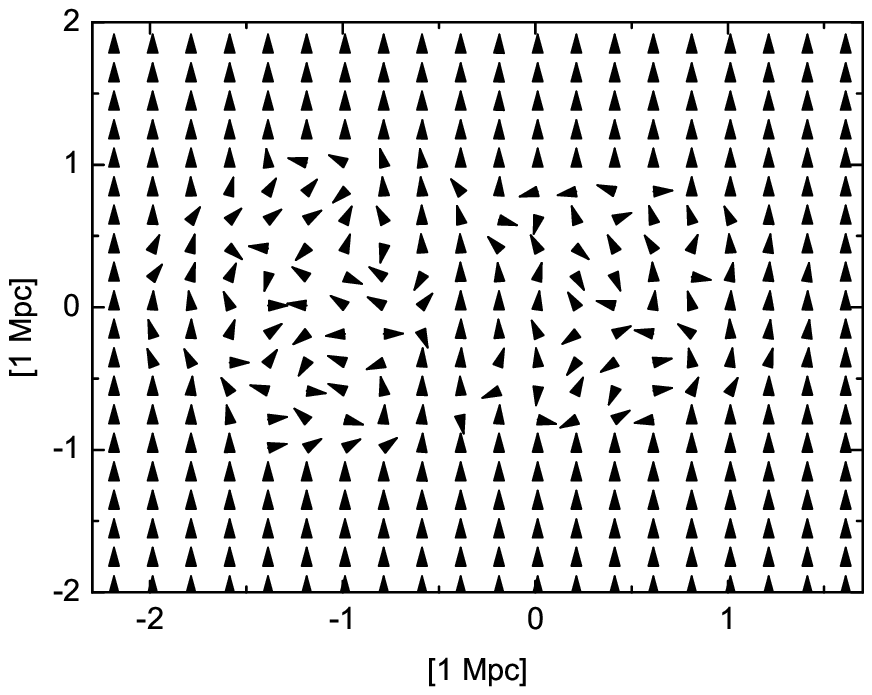} 
\caption{The problem of galaxy clusters collision. The density distribution in $cm^{-3}$ (top) and orientation of vector magnetic field (bottom) in the moment $t = 100 Myr$ are shown in figures. The grid with $512^3$ cells was used to compute the solution.}
\label{cluster}
\end{figure}

Within the problem of clusters collision, the interaction between two self-gravitating gaseous spheres in a weak vertical magnetic field was examined. To do this, on the distance 3 Mpc between centers of mass, the left gaseous sphere with mass $M_L = 10^{15} M_{\odot}$ and the right sphere with mass $M_R = 10^{14} M_{\odot}$ were set. The spheres have a temperature profile so that they are in gravitational equilibrium with their NFW density profiles. The speed of collision of each cluster is $v = 4000$ km s$^{-1}$. The value of the vertical magnetic field is 1 $\mu$G. The profile and orientation of vector of the magnetic field are shown in figure \ref{cluster}. 
Notice, that the result is qualitatively consistent with results in work \citep{Mastropietro_2008}. Also, significant reconstruction of the magnetic field due to clusters collision is visible. In our opinion, such problem is interesting in full formulation with consideration of collisionless component \citep{Kulikov_2014}. We are hoping that such formulation of the problem and the code we have, that able to solve it, will be interesting to researchers in the field of interaction of galaxy clusters and separate galaxies.

\subsection{Interaction between molecular cloud and interstellar medium}
\begin{figure}[ht]
\centering
\includegraphics[width=0.8\linewidth]{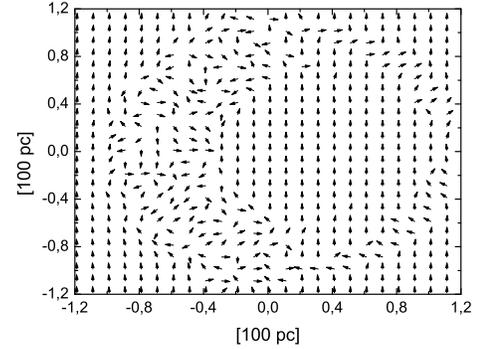} 
\caption{The problem of interaction between interstellar medium and molecular cloud. The orientation of vector magnatic field in the moment of time $t = 2.4$ Myr is in the figure.}
\label{cometmagfield}
\end{figure}

Within the problem of interaction between molecular cloud and interstellar medium we considered the model of hydrostatic equilibrium molecular cloud, and running on it rarefied ISM with speed $v = 45 km s^{-1}$. The size of the molecular cloud is $R = 100$ pc, and mass $10^7 M_{\odot}$. The profile of the density is:
$$
\rho(r) \sim 2r^3 - 3r^2 + 1
$$
the profile of the pressure is:
$$
p(r) \sim \pi \left( -\frac{r^8}{3} + \frac{44 r^7}{35} - \frac{6 r^6}{5} - \frac{4 r^5}{5} + 
\frac{8 r^4}{5} - \frac{2 r^2}{3} + \frac{1}{7} \right)
$$
The value of vertical magnetic field is $B_0$ = 0.05 $\mu$G. The results of simulation are in figures \ref{comet} and \ref{cometmagfield}.
In the beginning, there is an overruning of the flow and formation of the shock wave in front of the molecular cloud. Also, significant reconstruction of the magnetic field on the fronts of interaction is visible.

\subsection{Chemodynamics of evolution of MHD turbulence of the interstellar medium}
\begin{figure}[ht]
\centering
\includegraphics[width=0.8\linewidth]{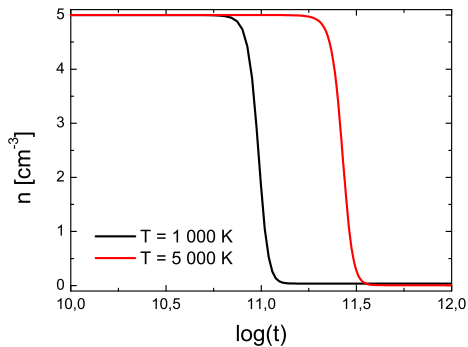} 
\caption{Behaviour of concentrations with temperature of region $T = 1000$ K and $T = 5000$ K. Rapid process of ionization within the time period $10^{11} < t < 10^{11.5}$ s for both temperatures is taking a place.}
\label{chemistryconc}
\end{figure}

The problem of chemodynamics of evolution of MHD turbulence of the interstellar medium was examined in full formulation with consideration of self-gravitation. To do this, the region $[256 pc]^3$ with vertical component of magnetic field, uniform initial concentration of atoms $n = 5 cm^{-3}$, initial random perturbation with speed $v_{rms} = 10 km/s$, initial value of plasma parameter $\beta_{th} = 8 \pi p_0 / B_0^2 = 25$, initial value of turbulent plasma parameter $\beta_{turb} = 8 \pi \rho v_{rms}^2 / B_0^2 = 25$, alfvenic Mach number $\mathcal{M} = 3.52$, was considered. The following eight reactions, that was also used in work \citep{Glover_2007}, was examined.
\begin{enumerate}
\item Molecular hydrogen formation \citep{Hollenbach_1979}: 
$$H + H + grain \rightarrow H_2 + grain$$
which held with speed $k_1$ and initiate heating of $\Gamma_1$.

\item Molecular hydrogen first dissociation \citep{Lepp_1983}:
$$H_2 + H \rightarrow 3H$$
which held with speed $k_2$ and initiate heating of $\Lambda_2$.

\item Molecular hydrogen second dissociation \citep{Martin_1998}:
$$H_2 + H_2 \rightarrow 2H + H_2$$
which held with speed $k_3$ and initiate cooling of $\Lambda_3$.

\item Molecular hydrogen photodissociation \citep{Glover_2007}:
$$H_2 + \gamma \rightarrow 2H$$
which held with speed $k_4$ and initiate heating of $\Gamma_4$.

\item Cosmic Ray ionization \citep{Glover_2007}:
$$H + c.r. \rightarrow H^+ + e$$
which held with speed $k_5$ and initiate heating of $\Gamma_5$.

\item Collision ionization \citep{Abel_1997}:
$$H + e \rightarrow H^+ + 2e$$
which held with speed $k_6$ and initiate cooling of $\Lambda_6$.

\item Radiative recombination \citep{Ferland_1992}:
$$H^+ + e \rightarrow H + \gamma$$
which held with speed $k_7$ and initiate cooling of $\Lambda_7$.

\item EI recombination on grains \citep{Weingartner_2001}:
$$H^+ + e + grain \rightarrow H + grain$$
which held with speed $k_8$ and initiate cooling of $\Lambda_8$.
\end{enumerate}
Each speed of reaction and also an analytical form of the cooling and heating functions are listed in the appendix. Effective adiabatic index was used in the following form:$$
\gamma = \frac{5 n_H +  5 n_e + 7 n_{H_2}}{3 n_H +  3 n_e + 5 n_{H_2}}
$$ 
Behaviour of concentration of different forms of hydrogen, which mostly was ionized, and molecular hydrogen was a several thousandth of a percent (see fig.  \ref{chemistryconc}), was modeled with using of code ChemPAK \citep{Chernykh_2009} for specific values of temperature $T = 1000$ K and $T = 5000$ K, and also for specific concentration of atomic neutral hydrogen. In numerical experiment concentrations behaved in a similar way.

The results of the simulation are shown in figure \ref{turbulencesim}. The formation of some small waves of density in the moments $t = 10$ and $t = 14$ Myr is visible in the figure. However, then the clusterization process is accelerated, that leads to the formation of clouds. Of course, we couldn't say is it possible to simulate the most known part of Carina Nebula, but, in our opinion, some kind of finger-like formations were obtained during the simulation.

Also dependence of the alfvenic speed on the gas density (see fig. \ref{turbulenceanalysis} on the left), and dependence of cosine of an angle of collinearity between velocity and vector of magnetic field on the gas density (see fig. \ref{turbulenceanalysis} on the right). It is clear from the figures, that for alfvenic Mach number the correlation $\mathcal{M} \sim n^2$ is traced (it is shown with white line), and the most part of the cloud $n > 10 cm^{-3}$ is in the over-alfvenic region (see fig. \ref{turbulenceanalysis} on the left). The reason of emergence of such mode is in magnetic turbulent interstellar medium in trans-alfvenic mode $\mathcal{M} \sim 1$ with $n \sim 1$. With such densities (see fig. \ref{turbulenceanalysis} on the right) contours of the cosine of an angle of collinearity between velocity and vector of magnetic field forms saddle-like structure, which means that the compression is along force lines of magnetic field. Then, further increase of mass and density of the cloud happening due to the influence of self-gravitation. In its turn, in dense clouds turbulence is just over-alfvenic with Mach number $\mathcal{M} > 100$. 
\section{Conclusion}
In the paper, the new code for numerical simulation of magnetic hydrodynamical astrophysical flows with consideration of chemical reactions is given. New original low-dissipation numerical method, based on a combination of operator-splitting approach and piecewise-parabolic method on a local stencil, for solving equations of magnetic-hydrodynamics is described in details. The scheme of program and results of scalability on classical multiprocessor architectures is given. Numerical method and its program implementation were verified with using of basic problems. Chemodynamics of hydrogen during the process of turbulent formation of molecular clouds was modeled.

This work is a part of the common joint ''Hydrodynamical Numerical Modelling of Astrophysical Flow at the Peta- and Exascale'', developed by our team at the Siberian Supercomputer Center ICMMG SB RAS.

\acknowledgments
The research work was supported by the Grant of the President of Russian Federation for the support of young scientists number MK -- 6648.2015.9, RFBR grants 15-31-20150, 15-01-00508, and 16-07-00434.

\appendix

\begin{figure}[ht]
\centering
\includegraphics[width = 0.4\linewidth]{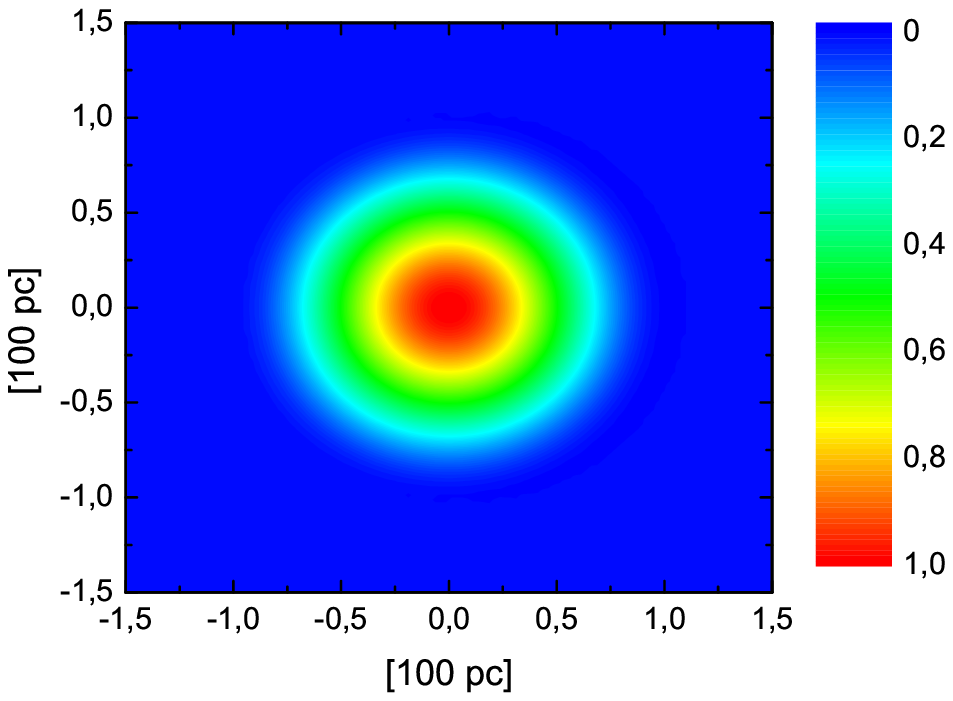} 
\includegraphics[width = 0.4\linewidth]{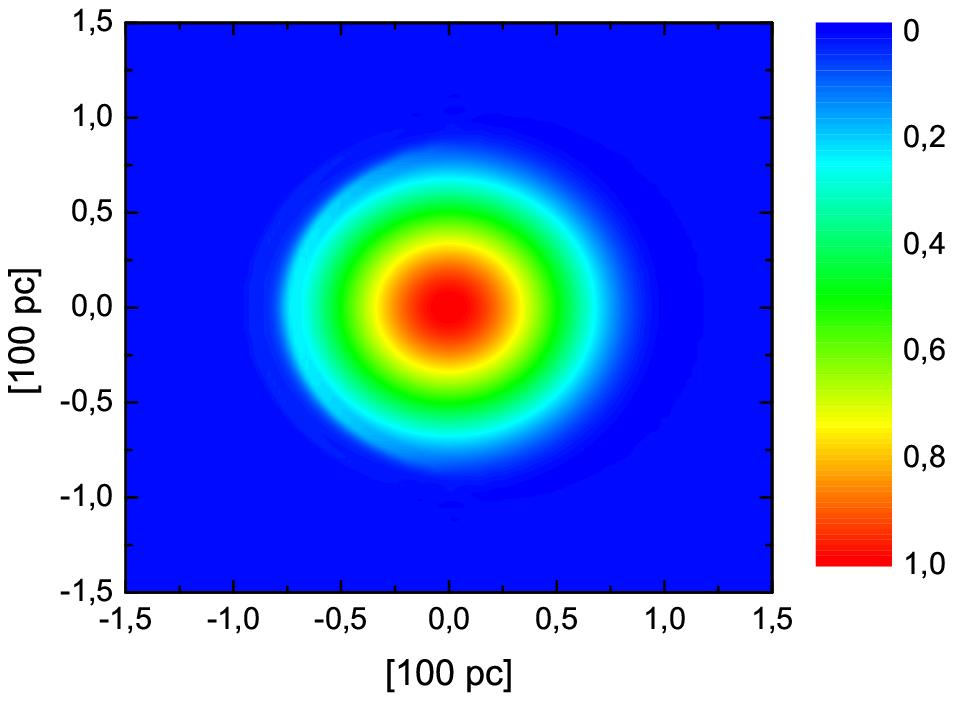} \\
\includegraphics[width = 0.4\linewidth]{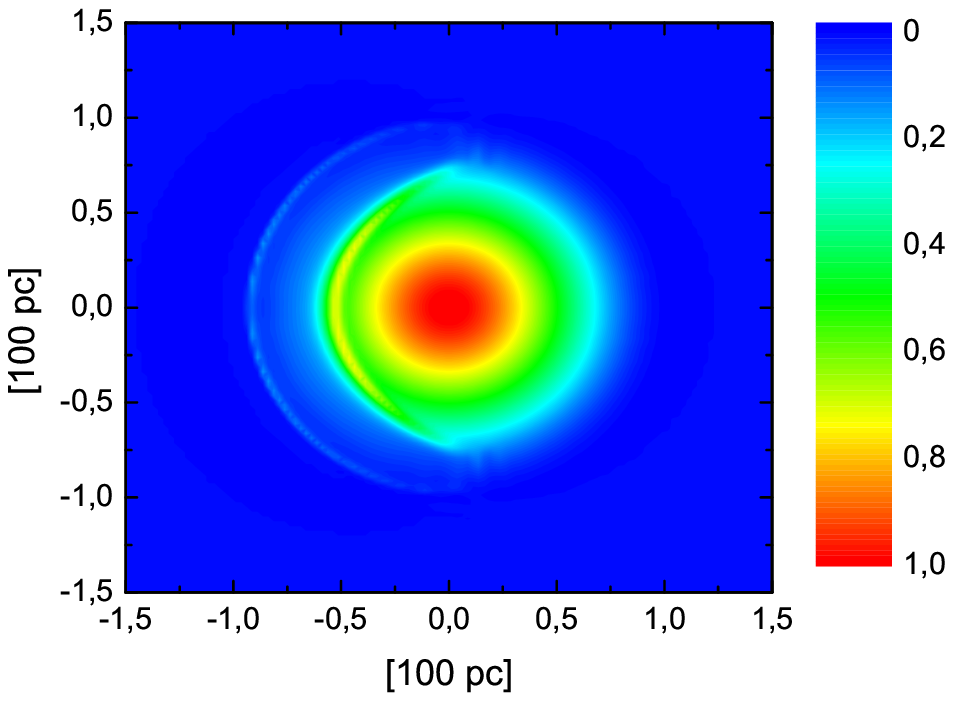} 
\includegraphics[width = 0.4\linewidth]{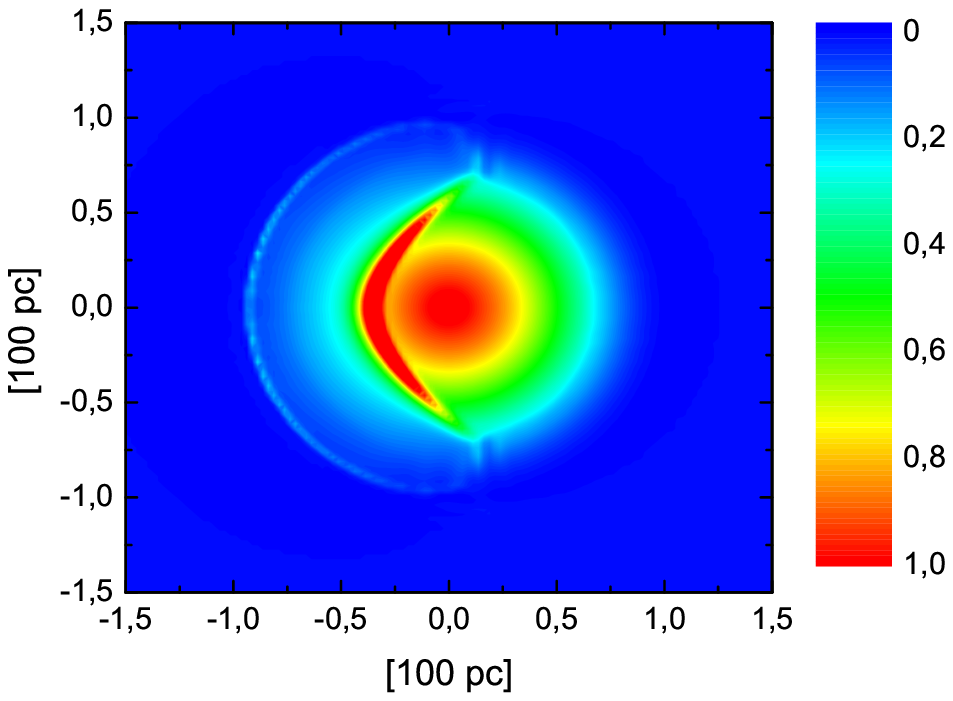} 
\caption{Problem of interaction between interstellar medium and molecular cloud. Distribution of density in $10^3 cm^{-3}$ in moments of time $t = 0.6$ Myr (top left), $t = 1.5$ Myr (top right), $t = 2.1$ Myr (bottom left), $t = 2.4$ Myr (bottom right). In the numerical experiment the grid with $512^3$ cells was used.}
\label{comet}
\end{figure}

\begin{figure}[ht]
\centering
\includegraphics[width = 0.32\linewidth]{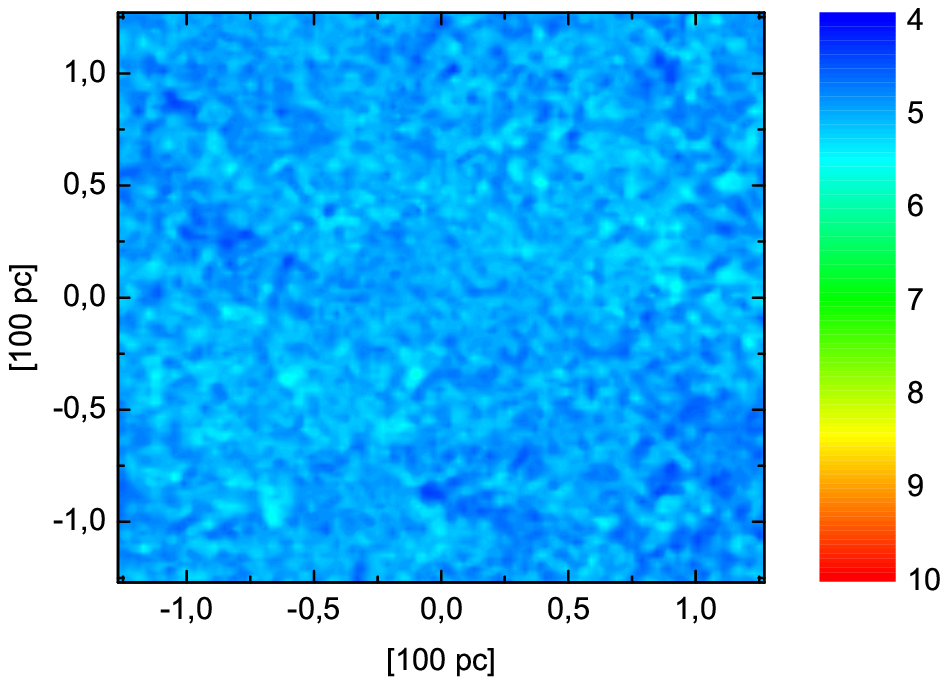} 
\includegraphics[width = 0.32\linewidth]{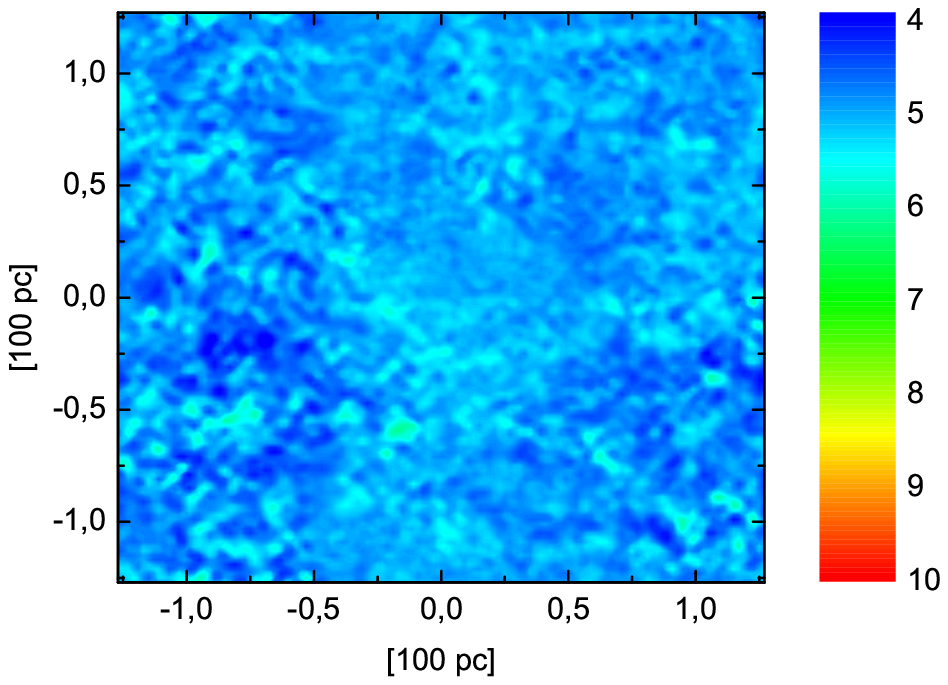} 
\includegraphics[width = 0.32\linewidth]{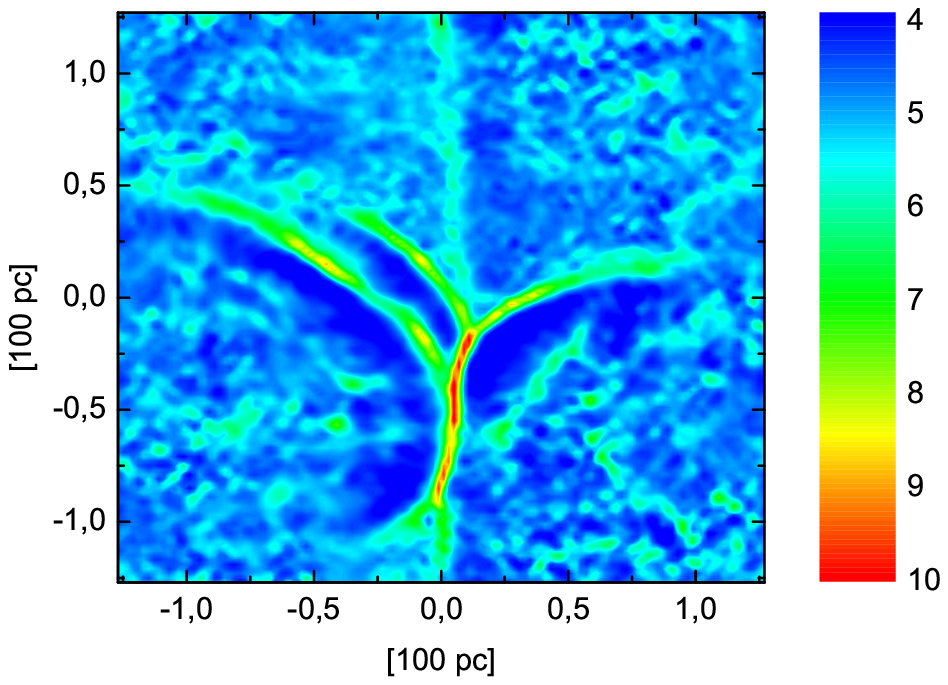} 
\caption{Problem of chemodynamics of evolution of MHD turbulence of interstellar medium. Concentration of the gas in $cm^{-3}$ in moments of time $t = 10$ Myr (left), $t = 14$ Myr (middle), $t = 15$ Myr (right) is given in figure. After the process of hydrogen ionization the cloud structures are formed. in the numerical experiment the grid with $512^3$ cells was used.}
\label{turbulencesim}
\end{figure}

\begin{figure}
\centering
\includegraphics[width = 0.4\linewidth]{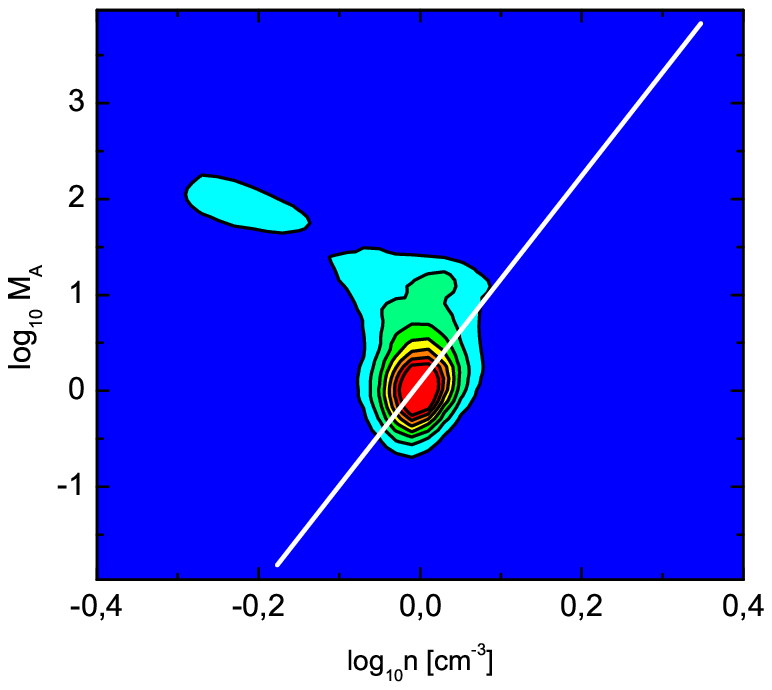}
\includegraphics[width = 0.4\linewidth]{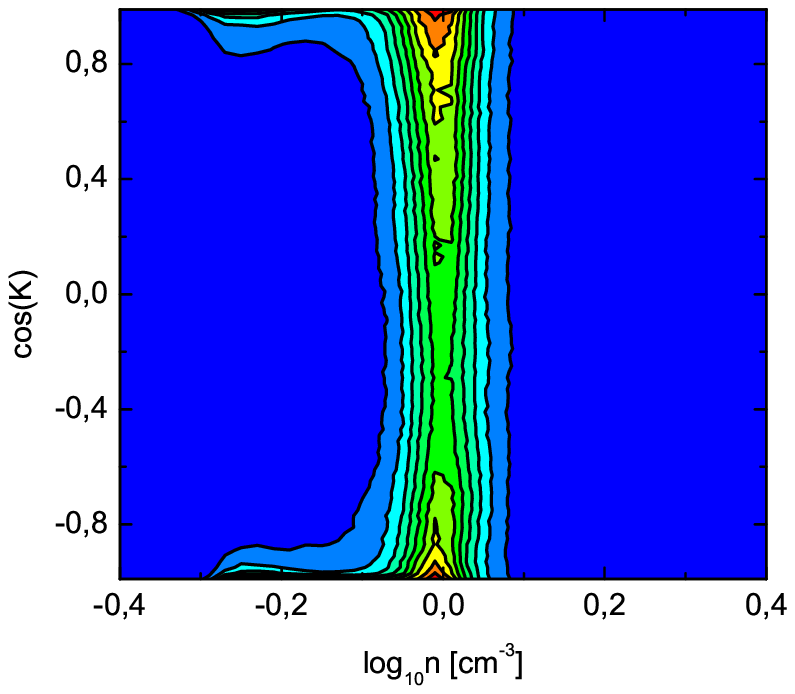}
\caption{Problem of chemodynamics of evolution of MHD turbulence of the interstellar medium. Dependence of alfvenic speed on gas density (left), and dependence of cosine of angle of collinearity between velocity and vector of magnetic field on gas density (right) are shown in figures.}
\label{turbulenceanalysis}
\end{figure}

Matrix of the right $R$ and left $L$ eigenvectors in the Riemann problem at the Eulerian stage are:
$$
R^T = \begin{pmatrix}
{{\alpha }_{f}}{{c}_{f}} & -{{\alpha }_{s}}{{c}_{s}}{{\beta }_{y}}sign\left( {{B}_{x}} \right) & -{{\alpha }_{s}}{{c}_{s}}{{\beta }_{z}}sign\left( {{B}_{x}} \right) & {{\alpha }_{s}}\sqrt{\rho }c{{\beta }_{y}}&{{\alpha }_{s}}\sqrt{\rho }c{{\beta }_{z}}&{{\alpha }_{f}}\gamma p  \\
-{{\alpha }_{f}}{{c}_{f}}&{{\alpha }_{s}}{{c}_{s}}{{\beta }_{y}}sign\left( {{B}_{x}} \right)&{{\alpha }_{s}}{{c}_{s}}{{\beta }_{z}}sign\left( {{B}_{x}} \right)&{{\alpha }_{s}}\sqrt{\rho }c{{\beta }_{y}}&{{\alpha }_{s}}\sqrt{\rho }c{{\beta }_{z}}&{{\alpha }_{f}}\gamma p \\
{{\alpha }_{s}}{{c}_{s}}&{{\alpha }_{f}}{{c}_{f}}{{\beta }_{y}}sign\left( {{B}_{x}} \right)&{{\alpha }_{f}}{{c}_{f}}{{\beta }_{z}}sign\left( {{B}_{x}} \right)&-{{\alpha }_{f}}\sqrt{\rho }c{{\beta }_{y}}&-{{\alpha }_{f}}\sqrt{\rho }c{{\beta }_{z}}&{{\alpha }_{s}}\gamma p \\
-{{\alpha }_{s}}{{c}_{s}}&-{{\alpha }_{f}}{{c}_{f}}{{\beta }_{y}}sign\left( {{B}_{x}} \right)&-{{\alpha }_{f}}{{c}_{f}}{{\beta }_{z}}sign\left( {{B}_{x}} \right)&-{{\alpha }_{f}}\sqrt{\rho }c{{\beta }_{y}}&-{{\alpha }_{f}}\sqrt{\rho }c{{\beta }_{z}}&{{\alpha }_{s}}\gamma p \\
0&-\frac{{{\beta }_{z}}}{\sqrt{2}}sign\left( {{B}_{x}} \right)&\frac{{{\beta }_{y}}}{\sqrt{2}}sign\left( {{B}_{x}} \right)&\sqrt{\frac{\rho }{2}}{{\beta }_{z}}&-\sqrt{\frac{\rho }{2}}{{\beta }_{y}}&0 \\
0&-\frac{{{\beta }_{z}}}{\sqrt{2}}sign\left( {{B}_{x}} \right)&\frac{{{\beta }_{y}}}{\sqrt{2}}sign\left( {{B}_{x}} \right)&-\sqrt{\frac{\rho }{2}}{{\beta }_{z}}&\sqrt{\frac{\rho }{2}}{{\beta }_{y}}&0
\end{pmatrix} 
$$
$$
L = \begin{pmatrix}
 \frac{{{\alpha }_{f}}{{c}_{f}}}{2{{c}^{2}}}&-\frac{{{\alpha }_{s}}{{c}_{s}}{{\beta }_{y}}}{2{{c}^{2}}}sign\left( {{B}_{x}} \right)&-\frac{{{\alpha }_{s}}{{c}_{s}}{{\beta }_{z}}}{2{{c}^{2}}}sign\left( {{B}_{x}} \right)&\frac{{{\alpha }_{s}}{{\beta }_{y}}}{2\sqrt{\rho }c}&\frac{{{\alpha }_{s}}{{\beta }_{z}}}{2\sqrt{\rho }c}&\frac{{{\alpha }_{f}}}{2\rho {{c}^{2}}} \\  
 -\frac{{{\alpha }_{f}}{{c}_{f}}}{2{{c}^{2}}}&\frac{{{\alpha }_{s}}{{c}_{s}}{{\beta }_{y}}}{2{{c}^{2}}}sign\left( {{B}_{x}} \right)&\frac{{{\alpha }_{s}}{{c}_{s}}{{\beta }_{z}}}{2{{c}^{2}}}sign\left( {{B}_{x}} \right)&\frac{{{\alpha }_{s}}{{\beta }_{y}}}{2\sqrt{\rho }c}&\frac{{{\alpha }_{s}}{{\beta }_{z}}}{2\sqrt{\rho }c}&\frac{{{\alpha }_{f}}}{2\rho {{c}^{2}}} \\  
 \frac{{{\alpha }_{s}}{{c}_{s}}}{2{{c}^{2}}}&\frac{{{\alpha }_{f}}{{c}_{f}}{{\beta }_{y}}}{2{{c}^{2}}}sign\left( {{B}_{x}} \right)&\frac{{{\alpha }_{f}}{{c}_{f}}{{\beta }_{z}}}{2{{c}^{2}}}sign\left( {{B}_{x}} \right)&-\frac{{{\alpha }_{f}}{{\beta }_{y}}}{2\sqrt{\rho }c}&-\frac{{{\alpha }_{f}}{{\beta }_{z}}}{2\sqrt{\rho }c}&\frac{{{\alpha }_{s}}}{2\rho {{c}^{2}}} \\  
 -\frac{{{\alpha }_{s}}{{c}_{s}}}{2{{c}^{2}}}&-\frac{{{\alpha }_{f}}{{c}_{f}}{{\beta }_{y}}}{2{{c}^{2}}}sign\left( {{B}_{x}} \right)&-\frac{{{\alpha }_{f}}{{c}_{f}}{{\beta }_{z}}}{2{{c}^{2}}}sign\left( {{B}_{x}} \right)&-\frac{{{\alpha }_{f}}{{\beta }_{y}}}{2\sqrt{\rho }c}&-\frac{{{\alpha }_{f}}{{\beta }_{z}}}{2\sqrt{\rho }c}&\frac{{{\alpha }_{s}}}{2\rho {{c}^{2}}} \\  
 0&-\frac{{{\beta }_{z}}}{\sqrt{2}}sign\left( {{B}_{x}} \right)&\frac{{{\beta }_{y}}}{\sqrt{2}}sign\left( {{B}_{x}} \right)&\frac{{{\beta }_{z}}}{\sqrt{2\rho }}&-\frac{{{\beta }_{y}}}{\sqrt{2\rho }}&0 \\  
 0&-\frac{{{\beta }_{z}}}{\sqrt{2}}sign\left( {{B}_{x}} \right)&\frac{{{\beta }_{y}}}{\sqrt{2}}sign\left( {{B}_{x}} \right)&-\frac{{{\beta }_{z}}}{\sqrt{2\rho }}&\frac{{{\beta }_{y}}}{\sqrt{2\rho }}&0 
\end{pmatrix}
$$

\appendix
Exact solution of the Riemann problem at the Eulerian stage for $x$ -- longitude component of velocity is written in following form:
$$
\mathbf{V_{x}} = 
\frac{\alpha_{s}^{2} c_{s}^{2} \left( v_{x}\left(c_{s} \tau \right) + v_{x}\left(- c_{s} \tau \right) \right)}{2 c^2}  +
\frac{\alpha_{f}   c_{f}   \alpha_{s}   c_{s}   \beta_{y}   sign \left( B_x \right) \left( v_{y}\left(c_{s} \tau \right) + v_{y}\left(- c_{s} \tau \right) - v_{y}\left(c_{f} \tau \right) - v_{y}\left(- c_{f} \tau \right) \right)}{2 c^2} + 
$$
$$
\frac{\alpha_{f}^{2} c_{f}^{2} \left( v_{x}\left(c_{f} \tau \right) + v_{x}\left(- c_{f} \tau \right) \right)}{2 c^2}  +
\frac{\alpha_{f}   c_{f}   \alpha_{s}   c_{s}   \beta_{z}   sign \left( B_x \right) \left( v_{z}\left(c_{s} \tau \right) + v_{z}\left(- c_{s} \tau \right) - v_{z}\left(c_{f} \tau \right) - v_{z}\left(- c_{f} \tau \right) \right)}{2 c^2} + 
$$
$$
\frac{\alpha_{f}   c_{f}   \alpha_{s}   \beta_{y} \left( b_{y}\left(- c_{f} \tau \right) - b_{y}\left(c_{f} \tau \right) \right)}{2 c \sqrt{\rho}} + 
\frac{\alpha_{f}   c_{f}   \alpha_{s}   \beta_{z} \left( b_{z}\left(- c_{f} \tau \right) - b_{z}\left(c_{f} \tau \right) \right)}{2 c \sqrt{\rho}}  +
\frac{\alpha_{f}^2   c_{f} \left( p\left(- c_{f} \tau \right) - p\left(c_{f} \tau \right) \right) }{2 \rho c^2}  +
$$
$$
\frac{\alpha_{f}   c_{s}   \alpha_{s}   \beta_{y} \left( b_{y}\left(- c_{s} \tau \right) - b_{y}\left(c_{s} \tau \right) \right)}{2 c \sqrt{\rho}}  + 
\frac{\alpha_{f}   c_{s}   \alpha_{s}   \beta_{z} \left( b_{z}\left(- c_{s} \tau \right) - b_{z}\left(c_{s} \tau \right) \right)}{2 c \sqrt{\rho}}  +
\frac{\alpha_{s}^2   c_{s} \left( p\left(- c_{s} \tau \right) - p\left(c_{s} \tau \right) \right) }{2 \rho c^2}  
$$
for $y$ -- lateral component of velocity (for another direction $z$ lateral directions change their places):
$$
\mathbf{V_{y}} = 
\frac{\alpha_{f}^{2}   c_{f}^{2}   \beta_{y}^{2} \left( v_{y}\left(c_{s} \tau \right) + v_{y}\left(- c_{s} \tau \right) \right)}{2 c^2}  + 
\frac{\alpha_{s}^{2}   c_{s}^{2}   \beta_{y}^{2} \left( v_{y}\left(c_{f} \tau \right) + v_{y}\left(- c_{f} \tau \right) \right)}{2 c^2}  +
\frac{\alpha_{s}^{2}  c_{s}^{2}   \beta_{y}   \beta_{z} \left( v_{z}\left(c_{f} \tau \right) + v_{z}\left(- c_{f} \tau \right) \right)}{2 c^2}  +
$$
$$ 
\frac{\alpha_{f}^{2}  c_{f}^{2}   \beta_{y}   \beta_{z} \left( v_{z}\left(c_{s} \tau \right) + v_{z}\left(- c_{s} \tau \right) \right)}{2 c^2}  +
\frac{sign \left( B_x \right)   \alpha_{f}   c_{f}   \beta_{y}   \alpha_{s}   c_{s} \left( v_{x}\left(c_{s} \tau \right) + v_{x}\left(- c_{s} \tau \right) - v_{x}\left(c_{f} \tau \right) - v_{x}\left(- c_{f} \tau \right) \right)}{2 c^2}  +
$$
$$
\frac{\beta_{z}^{2} \left(  v_{y}\left( c_{a} \tau \right) + v_{y}\left(- c_{a} \tau \right) \right)}{2} +
\frac{sign \left( B_x \right)   \alpha_{f}^{2}   c_{f}   \beta_{y}^{2} \left( b_{y}\left(c_{s} \tau \right) - b_{y}\left(- c_{s} \tau \right) \right)}{2 c \sqrt{\rho}}  + 
\frac{sign \left( B_x \right)   \alpha_{s}^{2}   c_{s}   \beta_{y}^{2} \left( b_{y}\left(c_{f} \tau \right) - b_{y}\left(- c_{f} \tau \right) \right)}{2 c \sqrt{\rho}}  + 
$$
$$
\frac{sign \left( B_x \right)   \beta_{z}^{2} \left( b_{y}\left( c_{a} \tau \right) - b_{y}\left(- c_{a} \tau \right) \right)}{2 \sqrt{\rho}}  +
\frac{sign \left( B_x \right)   \beta_{z} \beta_{y} \left( b_{z}\left(- c_{a} \tau \right) - b_{z}\left( c_{a} \tau \right) \right)}{2 \sqrt{\rho}} - 
\frac{\beta_{z} \beta_{y} \left( v_{z}\left( c_{a} \tau \right) + v_{z}\left(- c_{a} \tau \right) \right)}{2}  +
$$
$$
\frac{sign \left( B_x \right)   \alpha_{f}^{2}   c_{f}   \beta_{y}   \beta_{z} \left( b_{z}\left(c_{s} \tau \right) - b_{z}\left(- c_{s} \tau \right) \right)}{2 c \sqrt{\rho}}  + 
\frac{sign \left( B_x \right)   \alpha_{s}^{2}   c_{s}   \beta_{y}   \beta_{z} \left( b_{z}\left(c_{f} \tau \right) - b_{z}\left(- c_{f} \tau \right) \right)}{2 c \sqrt{\rho}}  +
$$
$$
\frac{sign \left( B_x \right)   \alpha_{f}   c_{f}   \beta_{y}   \alpha_{s} \left( p\left(- c_{s} \tau \right) + p\left(c_{f} \tau \right) - p\left(- c_{f} \tau \right) - p\left(c_{s} \tau \right) \right)}{2 \rho c^2} 
$$
for $y$ -- lateral component of the magnetic field (for another direction $z$ lateral directions changes their places):
$$
\mathbf{B_{y}} = 
\frac{\alpha_{f}^{2}  \beta_{y}^{2} \left( b_{y}\left(c_{s} \tau \right) + b_{y}\left(- c_{s} \tau \right) \right)}{2}  +
\frac{\alpha_{s}^{2}  \beta_{y}^{2} \left( b_{y}\left(c_{f} \tau \right) + b_{y}\left(- c_{f} \tau \right) \right)}{2}  +
\frac{\beta_{z}^{2} \left( b_{y}\left( c_{a} \tau \right) + b_{y}\left(- c_{a} \tau \right) \right)}{2} +
$$
$$
\frac{\alpha_{s}^{2}   \beta_{y}   \beta_{z} \left( b_{z}\left(c_{f} \tau \right) + b_{z}\left(- c_{f} \tau \right) \right)}{2} +
\frac{\alpha_{f}^{2}   \beta_{y}   \beta_{z} \left( b_{z}\left(c_{s} \tau \right) + b_{z}\left(- c_{s} \tau \right) \right)}{2} -
\frac{\beta_{z}   \beta_{y} \left( b_{z}\left( c_{a} \tau \right) + b_{z}\left(- c_{a} \tau \right) \right)}{2}  +
$$
$$
\sqrt{\rho} \frac{\alpha_{f}   \beta_{y}   \alpha_{s}   c_{s}   \left( v_{x}\left(c_{s} \tau \right) - v_{x}\left(- c_{s} \tau \right) \right) + \alpha_{s}   \beta_{y}   \alpha_{f}   c_{f}   \left( v_{x}\left(c_{f} \tau \right) - v_{x}\left(- c_{f} \tau \right) \right) + \alpha_{s}^{2}   \beta_{y}^{2}   c_{s}   sign \left( B_x \right)   \left( v_{y}\left(c_{f} \tau \right) - v_{y}\left(- c_{f} \tau \right) \right)}{2 c} + 
$$
$$
\frac{\beta_{z}^{2}  \sqrt{\rho}   sign \left( B_x \right) \left( v_{y}\left( c_{a} \tau \right) - v_{y}\left(- c_{a} \tau \right) \right) }{2}  + 
\frac{\alpha_{f}^{2}   \beta_{y}^{2}   c_{f}   sign \left( B_x \right) \left( v_{y}\left(c_{s} \tau \right) - v_{y}\left(- c_{s} \tau \right) \right)  \sqrt{\rho}}{2 c}  +
$$
$$
\frac{\beta_{z} \sqrt{\rho} \beta_{y} sign \left( B_x \right) \left( v_{z}\left(- c_{a} \tau \right) - v_{z}\left( c_{a} \tau \right) \right)}{2}  +
\frac{\alpha_{f}^{2}  \beta_{y}   c_{f}   \beta_{z}   sign \left( B_x \right)   \sqrt{\rho} \left( v_{z}\left(c_{s} \tau \right) - v_{z}\left(- c_{s} \tau \right) \right)}{2 c}  + 
$$
$$
\frac{\alpha_{s}^{2}  \beta_{y}   c_{s}   \beta_{z}   sign \left( B_x \right)   \sqrt{\rho} \left( v_{z}\left(c_{f} \tau \right) - v_{z}\left(- c_{f} \tau \right) \right)}{2 c} +
\frac{\alpha_{s}   \beta_{y}   \alpha_{f} \left( p\left(c_{f} \tau \right) + p\left(- c_{f} \tau \right) - p\left(c_{s} \tau \right) - p\left(- c_{s} \tau \right) \right)}{2 c \sqrt{\rho}} 
$$
for the pressure:
$$
\mathbf{P} = \frac{\rho   \alpha_{f}^{2}   c_{f} \left( v_{x}\left(- c_{f} \tau \right) - v_{x}\left(c_{f} \tau \right) \right)}{2} +
\frac{\sqrt{\rho}   \alpha_{f}   \alpha_{s}   \beta_{y} c \left( b_{y}\left(c_{f} \tau \right) + b_{y}\left(- c_{f} \tau \right) - b_{y}\left(c_{s} \tau \right) - b_{y}\left(- c_{s} \tau \right) \right)}{2} + 
$$
$$
\frac{\rho   \alpha_{s}^{2}   c_{s} \left( v_{x}\left(- c_{s} \tau \right) - v_{x}\left(c_{s} \tau \right) \right)}{2} +
\frac{\sqrt{\rho}   \alpha_{f}   \alpha_{s}   \beta_{z} c \left( b_{z}\left(c_{f} \tau \right) + b_{z}\left(- c_{f} \tau \right) - b_{z}\left(c_{s} \tau \right) - b_{z}\left(- c_{s} \tau \right) \right)}{2} +
$$
$$
\frac{\rho   \alpha_{f}   \alpha_{s}   c_{s}   \beta_{y}   sign \left( B_x \right) \left( v_{y}\left(c_{f} \tau \right) - v_{y}\left(- c_{f} \tau \right) \right)}{2}  + 
\frac{\rho   \alpha_{f}   \alpha_{s}   c_{f}   \beta_{y}   sign \left( B_x \right) \left( v_{y}\left(c_{s} \tau \right) - v_{y}\left(- c_{s} \tau \right) \right)}{2}  + 
\frac{\alpha_{f}^{2} \left( p\left(c_{f} \tau \right) + p\left(- c_{f} \tau \right) \right) }{2} +
$$
$$
\frac{\rho   \alpha_{f}   \alpha_{s}   c_{s}   \beta_{z}   sign \left( B_x \right) \left( v_{z}\left(c_{f} \tau \right) - v_{z}\left(- c_{f} \tau \right) \right)}{2}  + 
\frac{\rho   \alpha_{f}   \alpha_{s}   c_{f}   \beta_{z}   sign \left( B_x \right) \left( v_{z}\left(c_{s} \tau \right) - v_{z}\left(- c_{s} \tau \right) \right)}{2}  + 
\frac{\alpha_{s}^{2} \left( p\left(c_{s} \tau \right) + p\left(- c_{s} \tau \right) \right)}{2} 
$$

\appendix
The final equation for the electricity field:
$$
E_{x,i,k+1/2,l+1/2} = \frac{1}{4} \left(
\textbf{B}_{y,i,k,l+1/2} \textbf{U}_{z,i,k,l+1/2} - \textbf{B}_{z,i,k,l+1/2} \textbf{U}_{y,i,k,l+1/2} + \right.
$$
$$
\left.
\textbf{B}_{y,i,k+1,l+1/2} \textbf{U}_{z,i,k+1,l+1/2} - \textbf{B}_{z,i,k+1,l+1/2} \textbf{U}_{y,i,k+1,l+1/2} + 
\right.
$$
$$
\left.
\textbf{B}_{y,i,k+1/2,l+1} \textbf{U}_{z,i,k+1/2,l+1} - \textbf{B}_{z,i,k+1/2,l+1} \textbf{U}_{y,i,k+1/2,l+1} + 
\right.
$$
$$
\left.
\textbf{B}_{y,i,k+1/2,l} \textbf{U}_{z,i,k+1/2,l} - \textbf{B}_{z,i,k+1/2,l} \textbf{U}_{y,i,k+1/2,l} 
\right)
$$
$$
E_{y,i+1/2,k,l+1/2} = \frac{1}{4} \left(
\textbf{B}_{z,i,k,l+1/2} \textbf{U}_{x,i,k,l+1/2} - \textbf{B}_{x,i,k,l+1/2} \textbf{U}_{z,i,k,l+1/2} + \right.
$$
$$
\left.
\textbf{B}_{z,i+1,k,l+1/2} \textbf{U}_{x,i+1,k,l+1/2} - \textbf{B}_{x,i+1,k,l+1/2} \textbf{U}_{z,i+1,k,l+1/2} + 
\right.
$$
$$
\left.
\textbf{B}_{z,i+1/2,k,l} \textbf{U}_{x,i+1/2,k,l} - \textbf{B}_{x,i+1/2,k,l} \textbf{U}_{z,i+1/2,k,l} + 
\right.
$$
$$
\left.
\textbf{B}_{z,i+1/2,k,l+1} \textbf{U}_{x,i+1/2,k,l+1} - \textbf{B}_{x,i+1/2,k,l+1} \textbf{U}_{z,i+1/2,k,l+1}
\right)
$$
$$
E_{z,i+1/2,k+1/2,l} = \frac{1}{4} \left(
\textbf{B}_{x,i+1/2,k,l} \textbf{U}_{y,i+1/2,k,l} - \textbf{B}_{y,i+1/2,k,l} \textbf{U}_{x,i+1/2,k,l} + \right.
$$
$$
\left.
\textbf{B}_{x,i+1/2,k+1,l} \textbf{U}_{y,i+1/2,k+1,l} - \textbf{B}_{y,i+1/2,k+1,l} \textbf{U}_{x,i+1/2,k+1,l} +
\right.
$$
$$
\left.
\textbf{B}_{x,i,k+1/2,l} \textbf{U}_{y,i,k+1/2,l} - \textbf{B}_{y,i,k+1/2,l} \textbf{U}_{x,i,k+1/2,l} + 
\right.
$$
$$
\left.
\textbf{B}_{x,i+1,k+1/2,l} \textbf{U}_{y,i+1,k+1/2,l} - \textbf{B}_{y,i+1,k+1/2,l} \textbf{U}_{x,i+1,k+1/2,l} 
\right)
$$

\appendix
Speed of reactions, and also associated with them functions of cooling/heating has a following form. Speed of reactions Molecular hydrogen formation в ($cm^3 s^{-1}$) and further: 
$$
k_1 = \frac{3 \times 10^{-17} \sqrt{T/100} \times n/n_H}{1 + 0.4\sqrt{T/100} + 0.2(T/100) + 0.08 (T/100)^2} 
$$
heating function in ($ergs \times cm^{-3} s^{-1}$) and further:
$$
\Gamma_1 = 7.2 \times 10^{-12} \frac{n_{H_2}/n}{1 + \frac{4 - 0.416x - 0.327 x^2}{n}} 
$$
where $x = log(T/10^4)$. 

Speed of reaction for molecular hydrogen first dissociation:
$$
k_2 = 
\left\{
\begin{array}{@{\,}r@{\quad}l@{}}
6.11 \times 10^{-14} \exp \left(-2.93 \times 10^4 / T \right) & T > 7390 \\ 
2.67 \times 10^{-15} \exp \left(-(6750/T)^2 \right) & T \leq 7390
\end{array}\right.
$$
cooling function:
$$
\Lambda_2 = n_{H_2} \frac{L_H}{1 + L_H / L_L}
$$
where
$$
L_H = \left\{
\begin{array}{@{\,}r@{\quad}l@{}}
3.9 \times 10^{-19} \exp \left(-6118 / T \right) & T > 1087 \\ 
10^{-19.24+0.474x-1.247x^2} & T \leq 1087
\end{array}\right.
$$
$$
L_L = \left( n_{H_2}^{0.77} + 1.2 n_{H}^{0.77} \right) \times \left\{
\begin{array}{@{\,}r@{\quad}l@{}}
1.38 \times 10^{-22} \exp \left(-9243 / T \right) & T > 4031 \\ 
10^{-22.9-0.553x-1.148x^2} & T \leq 4031
\end{array}\right.
$$
where $x = log(T/10^4)$. 

Speed of reaction for molecular hydrogen second dissociation:
$$
k_3 = \left\{
\begin{array}{@{\,}r@{\quad}l@{}}
5.22 \times 10^{-14} \exp \left(-3.22 10^4/ T \right) & T > 7291 \\ 
3.17 \times 10^{-15} \exp \left(-(4060/T)-(7500/T)^2 \right) & T \leq 7291
\end{array}\right.
$$
cooling function
$$
\Lambda_3 = n_{H_2} \frac{L_H}{1 + L_H / L_L}
$$
where 
$$
L_H = 1.1 \times 10^{-13} \exp \left( -6744/T \right)
$$
$$
L_L = 8.18 \times 10^{-13} \left( n_H k_H + n_{H_2} k_{H_2} \right)
$$
where
$$
k_{H_2} = 6.29 \times 10^{-15} \times 1.38 \times f(T) / f(4500)
$$
where $f(T) = \sqrt{T} \alpha \exp{\alpha}$, $\alpha = 1 + (kT)^{-1}$, $k$ -- Boltzmann constant.

Speed of reaction for molecular hydrogen photodissociation:
$$
k_4 = \xi_{diss}(0) f_{shield}(N_{H_2}) f_{dust}(A_{V})
$$ 
where $\xi_{diss}(0) = 3.3 \times 1.7 \times 10^{-11}$ -- unshielded photodissociation rate \citep{Draine_1978}, $f_ {dust} (A_ {V}) =exp (-\tau_ { d, 1000 } (A_V)) $ -- absorption rate on dust \citep{Draine_1996}, where $\tau_{d,1000}(A_V) = 3.74 A_{V} = 10^{-21}\left( N_{H} + N_{H_2} \right)$ -- optical depth on dust particles on wavelength $\lambda = $ \AA 1000, where $N_{H}$ and $N_{H_2}$ -- column density. The function of the coefficient of self-shielding can be approximated \citep{Draine_1996}:
$$
f_{shield}(N_{H_2}) = \frac{0.965}{(1 + x/b_5)^2} + \frac{0.035}{\sqrt{1+x}}exp\left( -8.5 \times 10^{-4} \sqrt{1+x} \right)
$$
where $x = N_{H_2}/5 \times 10^ {10}$ m$^2$, $b_5 = b/10^7$ m/s, where $b$ -- the parameter of Doppler expansion, heating function
$$
\Gamma_4 = 6.4 \times 10^{-13} \times k_4 \ n_{H_2}
$$

Speed of reaction for Cosmic Ray ionization $k_5 = 6 \times 10^{-18} n_{H_2}$, heating function
$\Gamma_5 = 1.92 \times 10^{-28} n$.

Speed of reaction for collision ionization:
$$
k_6 = \exp \left( -32.7 + 13.5 \log(T) - 5.7 \log^2(T) + 1.5 \log^3(T) - 0.3 \log^4(T) \right.
$$
$$
\left. + 3.4(-2) \log^5(T) - 2.6(-3) \log^6(T) + 1.1(-4) \log^7(T) - 2.1(-6) \log^8(T) \right)
$$
cooling function
$$
\Lambda_6 = 2.18 \times 10^{-11} k_6
$$

SPeed of reaction for Radiative recombination:
$$
k_7 = 10^{\frac{-10.78+4.68x -0.87x^2+0.08x^3-3.87(-3)x^4}{1 -0.38x+0.06x^2-5.1(-3)x^3+2.4(-4)x^4}}
$$ 
where function $x = log(T)$ and cooling function 
$$
\Lambda_7 = 4.65 \times 10^{-30} \times T^{0.94} \times \left( \frac{\exp(-0.75 \times 10^{-20} (N_H + N_{H_2})) \sqrt{t}}{n_e} \right)^{0.74/T^{0.068}}
$$

Speed of reaction for EI recombination on grains:
$$
k_8 = \frac{12.25 \times 10^{-14}}{1 + 8.074(-6) \times 10^{2.756} \left( 1 + 5.087(2) \times T^{1.586(-2)} 10^{-1.8892 - 4.4(-5) log(T)} \right)}
$$
cooling function
$$
\Lambda_8 = 5.7 \times 10^{-26} \times \left( T/10^4 \right)^{0.8}
$$

\end{document}